\documentclass[preprint,12pt,3p]{elsarticle}
\usepackage[T1]{fontenc}
\usepackage{times}
\usepackage{graphics,graphicx,subcaption}
\usepackage{enumerate,epsfig}
\usepackage{multirow,afterpage,rotating}
\usepackage{amsfonts}
\usepackage{amsmath,amssymb}
\usepackage{caption}
\usepackage{soul}
\usepackage{url}
\usepackage{inputenc}
\usepackage[misc,geometry]{ifsym}
\usepackage[colorlinks=true]{hyperref}
\usepackage{dcolumn}
\usepackage{wrapfig,dsfont}
\usepackage{mathtools}
\usepackage{color}
\usepackage{tikz}
\usepackage{float}

\DeclareCaptionFont{red}{\color{red}}

\newcommand{\solidcircle}[1][red]{\tikz{\draw[black,fill=#1] (0,0) circle (.5ex);}}
\newcommand{\solidsquare}[1][red]{\tikz{\draw[black,fill=#1] (0,0) rectangle (1ex,1ex);}}

\definecolor{henna}{rgb}{0.8,0.8,0.1}
\definecolor{skyblue}{rgb}{0.1,0.8,0.8}
\definecolor{purple}{rgb}{0.8,0.1,0.8}
\definecolor{darkblue}{rgb}{0.1,0.1,0.8}
\definecolor{darkgreen}{rgb}{0.1,0.8,0.1}
\definecolor{darkred}{rgb}{0.8,0.1,0.1}

\biboptions{numbers,sort&compress}

\hypersetup{urlcolor=blue, citecolor=red}

\newlength{\defbaselineskip}
\newcommand{\setlinespacing}[1]%
           {\setlength{\baselineskip}{#1 \defbaselineskip}}

\journal{Soft Computing, Springer}
\begin{document}

\begin{frontmatter}

\title{Convolutional Neural Network-Bagged Decision Tree: A hybrid approach to reduce electric vehicle's driver's range anxiety by estimating energy consumption in real-time}

\author[label1]{Shatrughan Modi\corref{cor1}}
\address[label1]{Computer Science and Engineering Department, Thapar Institute of Engineering and Technology, India}

\cortext[cor1]{corresponding author}

\ead{shatrughanmodi@gmail.com}

\author[label1]{Jhilik Bhattacharya}
\ead{jhilik@thapar.edu}

\author[label5]{Prasenjit Basak}
\address[label5]{Electrical and Instrumentation Engineering Department, Thapar Institute of Engineering and Technology, India}
\ead{prasenjit@thapar.edu}

\begin{abstract}
    To overcome range anxiety problem of Electric Vehicles (EVs), an accurate real-time energy consumption estimation is necessary, which can be used to provide the EV's driver with information about the remaining range in real-time. A hybrid CNN-BDT approach has been developed, in which Convolutional Neural Network (CNN) is used to provide an energy consumption estimate considering the effect of temperature, wind speed, battery's SOC, auxiliary loads, road elevation, vehicle speed and acceleration. Further, Bagged Decision Tree (BDT) is used to fine tune the estimate. Unlike existing techniques, the proposed approach doesn't require internal vehicle parameters from manufacturer and can easily learn complex patterns even from noisy data. Comparison results with existing techniques show that the developed approach provides better estimates with least mean absolute energy deviation of 0.14.
\end{abstract}

\end{frontmatter}

\section{Introduction}\label{Sec:Introduction}

\begin{table*}[h!]
    \centering
    \label{Tab:Nomenclature}
    \begin{tabular}{|ll|}
        \hline
        \textbf{Abbreviations} &                                \\
        $EV$        &   Electric Vehicle                        \\
        $CNN$       &   Convolutional Neural Network            \\
        $SOC$       &   State of Charge                         \\
        $BDT$       &   Bagged Decision Tree                    \\
        $PCE$       &   Power Consumption Estimation            \\
        $RMSE$      &   Root Mean Square Error                  \\
        $MAE$       &   Mean Absolute Error                     \\
        $Corr$      &   Correlation                             \\
        $MAE_{dev}$ &   Mean Absolute Energy Deviation          \\
        $MPTDC$     &   Mean Prediction Time per Drive Cycle    \\
        $UDDS$      &   Urban Dynamometer Driving Schedule      \\
        $SFTP$      &   Supplemental Federal Test Procedures    \\
        $FASTSim$   &   Future Automotive Systems Technology Simulator  \\

        \textbf{Symbols} &                                      \\
        $veh_{sp}$  &   Vehicle's Speed                         \\
        $road_{el}$ &   Road Elevation                          \\
        $veh_{acc}$ &   Vehicle's Acceleration                  \\
        $aux_{ld}$  &   Auxiliary Loads                         \\
        $wind_{sp}$ &   Wind Speed                              \\
        $batt_{soc}$&   State of charge of battery              \\
        $env_{temp}$&   Environmental Temperature               \\
        $SOC_i$     &   Battery's state of charge at $i^{th}$ time instant          \\
        $E_{cap}$   &   Battery's rated energy capacity                             \\
        $Est_{pow}$ &   Power consumption estimated by the proposed approach        \\
        $Act_{pow}$ &   Actual power consumption as given in dataset                \\
        $\overline{Est_{pow}}$ &   Mean of estimated power consumption              \\
        $\overline{Act_{pow}}$ &   Mean of actual power consumption                 \\
        $P_{reg}$   &   Regenerative power                      \\
        $\eta_{te}$ &   Transmission efficiency                 \\
        $\delta$    &   Driving efficiency                      \\
        $m$         &   EV's weight related coefficient         \\
        $\rho$      &   Air density                             \\
        $C_D$       &   Aerodynamic drag coefficient            \\
        $P_{accessory}$        &    Power consumed by accessories                   \\
        $\eta_m$    &   Motor efficiency                        \\
        $k$         &   Percentage of energy restored by the motor during braking   \\

        \hline
    \end{tabular}
    
\end{table*}

Transportation industry is seeing Electric Vehicles (EVs) as the future mode of transport due to depleting fossil fuels and increasing need of overcoming global warming due to vehicle pollution. For this, it is necessary to overcome the barriers faced by the EVs for their mass adoption. Number of studies \cite{ZHANG2018527, WOLFF2019262, FENG2013135, CARLEY201339, HACKBARTH20135} have been conducted to analyse the major factors which influence the market penetration of electric vehicles. It has been found that short driving range, scarcity of public charging infrastructure, long recharging time and high initial cost are the major factors among others. Consumers show more interest in buying a plug-in hybrid vehicle than all electric battery vehicle. The main reason for this is short driving range of EVs which makes the driver anxious that whether he/she will be able to reach his/her destination or not with current state of charge of battery. Although, advancements in battery technology \cite{MANZETTI20151004, Cano2018} helped in increasing the consumer's confidence but reliable energy usage estimate in real time is the key that can boost the consumer's trust by a great deal. Hence, the main focus of this work is to develop a reliable methodology for accurate energy consumption estimation of EVs.

There are number of methodologies developed by researchers to estimate EVs energy consumption. These can be categorised based on the granularity level (i.e. microscopic, mesoscopic and macroscopic), at which they were developed for different applications. The macroscopic models \cite{WU201552, 6861542, 7313117, LIU201774, LIU2018324, FETENE20171}, are those which can be used to estimate EVs total energy consumption at the trip level which include number of connecting roads with different traffic conditions. The mesoscopic models \cite{QI201836, de2017data, 7007121, Yao2013}, are the models used to estimate the EVs energy consumption for each road in the road network so that energy consumption cost can be assigned to each road in the road network and then can be used to plan the optimal route. The microscopic models \cite{ZHANG2015177, YANG201441, GALVIN2017234, FIORI2016257, MODI2019}, are the models which provide more detailed second-by-second energy consumption for a EV based on the different influencing factors. The microscopic models can be aggregated to get the energy consumption estimates for each road in the road network and hence can serve the purpose of mesoscopic models and then can be further aggregated to get the energy consumption estimate for the whole trip and serve the purpose of macroscopic models. Also, these models can be used for providing the drivers with real-time instruction based on the current energy consumption and remaining energy in the battery to maintain an optimal speed or to take an alternative path so that the driver can reach his/her destination with minimum energy consumption. Hence, these models can help in reducing driver's range anxiety. Due to this, in this work a microscopic model which can give real time energy consumption estimates has been developed.

The approaches developed so far for EVs energy consumption estimation have either used simulation techniques \cite{GENIKOMSAKIS201798, 7902061, 4168023} or regression based techniques like linear regression \cite{GALVIN2017234, LIU20161351, FETENE20171}, polynomial regression \cite{WU201552, de2017data, LIU2018324, QI201836, YANG201441}, logarithmic regression \cite{LIU201774}. A few Neural Network (NN) \cite{6861542, 7313117}, Neuro Fuzzy \cite{1005395, DAI2015350} and Convolutional Neural Network (CNN) \cite{MODI2019} based techniques were also developed. The simulation based techniques require vehicle specific calibration, which require internal vehicle parameters, like efficiency curve of the motor and internal resistance of the battery etc. These vehicle specific parameters are very hard to obtain because vehicle manufacturers doesn't share this information in public domain. Also, due to vehicle specific calibration the simulation based models can not be generalized. On the other hand, real world data is required for the regression based techniques and most of these techniques are sensitive to noise \cite{kalapanidas2003machine}, which is generally the case with real-world data, as it is mostly obtained from different sensors. Although, the techniques based on NN \cite{6861542, 7313117, MODI2019} can handle the noisy data better than other regression techniques but they lack the applicability in real-world. The models presented in \cite{6861542, 7313117}, provide single energy consumption output for the whole trip. Hence, these models can not be used by the drivers for real time guidance based on the current energy consumption to maintain an optimal speed or to take an alternative path, so that the driver can reach his/her destination with minimum energy consumption. Also, these models have considered speed, acceleration, jerk and road related parameters but there are many other influencing parameters that need to be considered. The neuro-fuzzy based techniques \cite{1005395, DAI2015350}, provide encouraging results but they can be improved further for representing the non-linear patterns more accurately by following techniques presented in \cite{AbuArqub2016, Arqub2017, AbuArqub2017, Arqub2020}. Similarly, the CNN model developed in \cite{MODI2019}, show encouraging results with high robustness and considered road elevation, vehicle speed and tractive effort as input. There are number of studies \cite{Sweeting2011, Yi2017} which show that the energy consumption of EVs get influenced by several other parameters (like environment temperature, wind, battery's State of Charge (SOC) and auxiliary loads etc.) significantly. So, to get accurate estimates it is important to consider the effect of all these factors.

To address the research gaps discussed above, a hybrid approach using Convolutional Neural Network (CNN) and Bagged Decision Tree (BDT) has been developed for providing a real-time accurate estimate of EVs energy consumption. The CNN used in current work is a multi-channel CNN i.e. there are multiple parallel branches where each branch extract important features from individual input parameter and then extracted features are combined for predicting the energy consumption. The proposed approach also uses a BDT, which consist of multiple decision trees ensembled together. The BDT is used as a fine tuner to improve the initial estimate given by CNN by decreasing the prediction error further. As discussed above, the one of main challenge is the requirement of vehicle specific parameters from the vehicle manufacturers to calibrate the simulation models. As the vehicle manufacturers do not provide these vehicle specific parameters openly, it is very hard to calibrate the simulation models without these. The proposed CNN-BDT model does not require any vehicle specific parameters for its working. It takes seven input parameters namely, vehicle speed, vehicle acceleration, road elevation, wind speed, auxiliary loads, environmental temperature and initial battery's SOC. These parameters are easily available, for instance, vehicle speed is easily available from vehicle, vehicle acceleration can be computed from vehicle speed, Geographic Information System (GIS) can be used for obtaining road elevation, freely available weather API's (like API's from OpenWeather \cite{OpenWeather}) can be used to obtain wind speed and environmental temperature etc. Also one of the problem with regression techniques is their noise sensitivity. The NN based techniques perform much better than other regression techniques because the NN based techniques can recognize the patterns more easily even from noisy data. The proposed technique is a hybrid technique, which uses CNN as one of its sub-module, hence can handle the noisy data quite easily. In comparison to the other NN based techniques \cite{6861542, 7313117}, which do not provide real-time output, the proposed CNN-BDT technique provide energy consumption output in real-time. Hence, the proposed CNN-BDT technique can be used for real-time driver guidance for optimal battery usage. The current proposed technique is an extension to the technique presented in \cite{MODI2019} and can provide better real-time energy consumption estimates in terms of accuracy, as it considers the effect of almost all the influencing factors namely, vehicle speed, acceleration, wind speed, auxiliary loads, battery's SOC, environmental temperature and road elevation. The major contributions of the current work are as discussed below:

\begin{enumerate}[i)]
    \item A hybrid CNN-BDT approach has been developed to provide the EV's driver with reliable real-time estimates of energy consumption. The approach can also be used to provide trip level energy consumption estimates.
    \item The proposed technique uses a multichannel CNN model which takes multiple one-dimensional time series inputs and extract features from them and then predict the energy consumption estimate based on the extracted features. Thus, the approach is suitable for modelling the non-linear relationship among the influencing parameters and can generalize well even in the presence of sensor noise or outlier data.
    \item Bagged Decision Tree (BDT) is used as a fine tuner to fine tune the estimate provided by CNN based on the current input parameters. Hence, the proposed technique uses the strengths of both CNN and BDT and provide accurate results.
    \item The proposed technique takes seven input parameters namely, vehicle speed, acceleration, wind speed, auxiliary loads, battery's SOC, environmental temperature and road elevation. So, it takes into account the influence of all the major factors.
    \item The proposed technique doesn't require any of the internal vehicle parameters from vehicle manufacturer and hence, can be easily trained for any other vehicle.
\end{enumerate}

The paper has been further organized as follow. Section \ref{Sec:ArchitectureDescription} provide the detailed description of the proposed methodology. Section \ref{Sec:ExperimentalSetup} discusses the experimental setup which includes datasets used, preprocessing of data and the hyperparameters used. The results obtained from the experiments and comparison results of the proposed technique with existing techniques have been discussed in Section \ref{Sec:Results} and Section \ref{Sec:Conclusion} concludes the paper.

\section{Proposed Architecture}\label{Sec:ArchitectureDescription}
\begin{figure}[!ht]
      \centering
      \includegraphics[width=\linewidth]{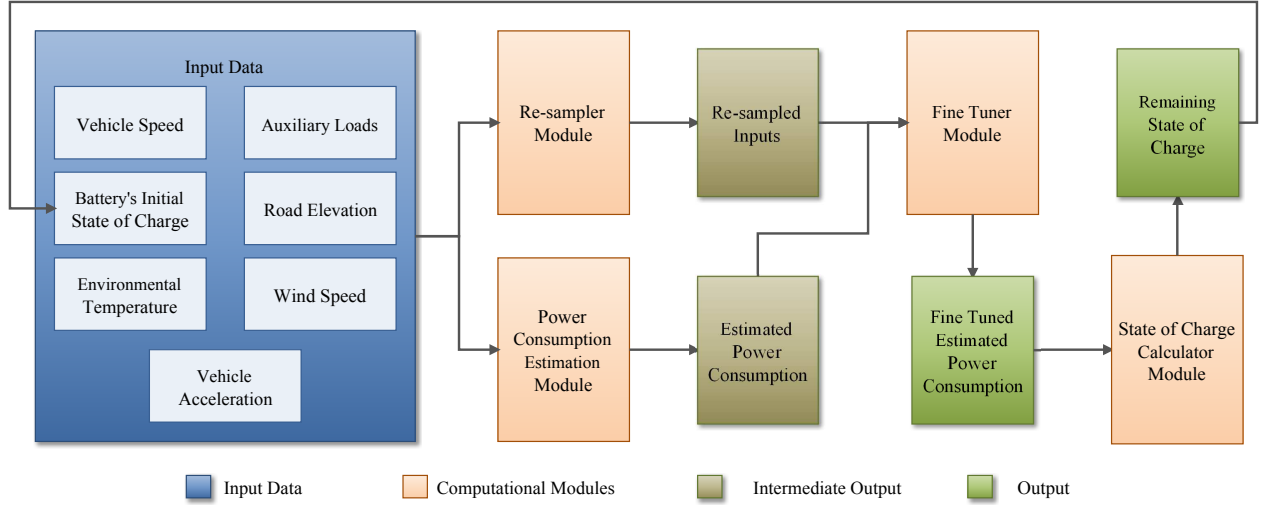}
      \caption{Architecture of Proposed Approach}\label{Fig:ArchitectureDiagram}
\end{figure}

There are number of factors like environment temperature, wind speed, road elevation, battery's state of charge (SOC) etc which influence the energy consumption of an EV. These factors vary a lot in real life and have a non-linear relationship among them. So, to effectively represent their non-linear relationship a hybrid CNN-BDT approach has been adopted, which can provide accurate power / energy consumption estimates of an EV under different conditions. Figure \ref{Fig:ArchitectureDiagram}, shows the proposed approach's architecture. There are four main computational modules of the proposed approach namely, Power Consumption Estimation (PCE) Module, Re-sampler Module, Fine Tuner Module and State of Charge Calculator Module. The proposed approach takes seven inputs namely, vehicle speed, vehicle acceleration, environment temperature, wind speed, auxiliary loads, road elevation and battery's initial SOC. As discussed in Section \ref{SubSec:DataSets}, these inputs are small partitioned time series of 10 Hz frequency each of 10 sec duration i.e. each partition has 100 readings for 10 sec duration. The environmental temperature and battery's initial SOC does not change much in the duration of 10 sec so they have been considered as constant for a particular partition. The inputs are provided simultaneously to the PCE and Re-sampler module. The PCE module consist of a CNN model, which uses these inputs and provide an estimated power consumption by the EV. The power consumption estimate given by PCE module is of 1 Hz frequency. The Re-sampler module takes the seven inputs and up/down sample them to match the frequency of inputs with output from PCE module i.e. 1 Hz. Then, the re-sampled inputs along with estimated power consumption are further passed through a Fine Tuner module which uses BDT and fine tune the estimate of power consumption by reducing the error. The fine tuned estimated power consumption is then used by the State of Charge Calculator module. It calculates the remaining state of charge of the battery based on the power consumed by the EV. This calculated remaining SOC is then used as the initial battery SOC for the next partition of 10 sec. Following sub-sections discuss in detail the working of these four computational modules of proposed architecture.

\subsection{Power Consumption Estimation (PCE) Module}\label{SubSec:PowerConsumptionEstimationModule}
\begin{figure}[!ht]
      \centering
      \includegraphics[width=\linewidth]{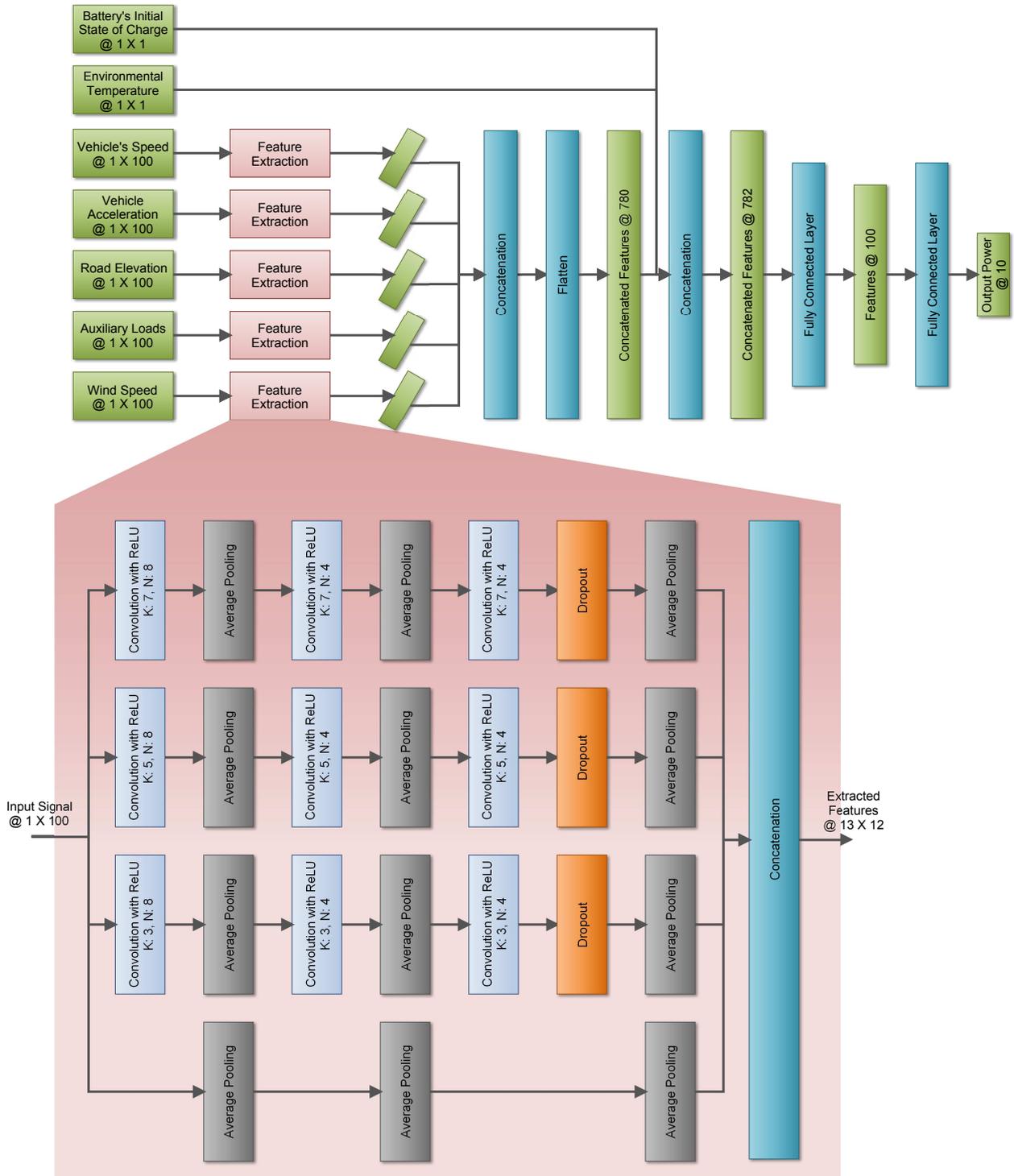}
      \caption{Architecture of Power Consumption Estimation (PCE) Module. (Numbers after K, N and @ represent the kernel size, number of kernels and the dimension of vector, respectively)}\label{Fig:CNNArchitecture}
\end{figure}

The Power Consumption Estimation (PCE) Module is the module responsible for estimating the power consumption of EV under different environmental and road conditions. For this a multi-channel CNN, as shown in Figure \ref{Fig:CNNArchitecture}, has been developed. The multi-channel CNN is inspired from the network architecture used in \cite{8373818}, originally developed for hand gesture classification from time series pose data. In this work, the proposed architecture takes seven input time series namely, vehicle's speed ($veh_{sp}$), road elevation ($road_{el}$), vehicle's acceleration ($veh_{acc}$), auxiliary loads ($aux_{ld}$), wind speed ($wind_{sp}$), initial state of charge of battery ($batt_{soc}$) and environmental temperature ($env_{temp}$). As discussed in Section \ref{SubSec:DataSets}, all of these time series (say $x$) in the dataset were recorded at 10 Hz frequency and were partitioned into $n$ small time series (say $x^i , i = 1,2,...,n$) each of 10 seconds duration. These small time series each of 10 sec duration became the input for the network. It has been observed that out of these seven input parameters first five parameters vary a lot but the environmental temperature ($env_{temp}$) and battery's initial SOC ($batt_{soc}$) does not change during the interval of 10 seconds. Due to this, these two parameters have been considered as constant for each partition and no feature extraction has been performed for these two parameters. Each of the other five parameters have been passed to a separate feature extraction module, each of them have four separate branches to extract features. Each feature extraction module has one residual branch and three similar convolutional branches.

Residual branches make gradient backpropagation better during the training and hence the network optimize easily which ultimately increase the accuracy of network \cite{7780459}. The residual branch is acting like an identity function but instead of giving same output as input three average pooling layers have been used. The average pooling layers downsample the data by taking the average of the input data from a particular region. The pooling layers make the CNN locally invariant i.e. the CNN can extract the same features from the input regardless of rotation, scaling or shifting of features \cite{s17040818}. So, the pooling layers not only reduce the network scale, but also extract the important features from the input. This also helps in avoiding overfitting of the network. The main feature extraction is done by the three convolutional branches which have similar architecture, as discussed below.

Each convolutional branch has three convolutional layers followed by average pooling layer except the last convolution layer which is followed by a dropout layer and then a pooling layer. Convolutional layers are different from traditional fully connected layers as they apply convolutional filters on the input. The convolutional filters extract the local features by focusing on a particular region of the input. One convolutional layer can have hundred of such filters and extract hundreds of features from the input. This helps the convolutional layers to learn the complex patterns. In a single branch, the convolutional layers differs from each other in terms of number of kernels used, for instance, in each convolutional branch the first convolutional layer has 8 kernels whereas the second and third convolutional layer has 4 kernels. Convolutional branches differ from each other in terms of size of kernels i.e. one convolutional branch has kernels of size 3, second has kernels of size 5 and third has kernels of size 7. Branches with different kernel size help the network to extract and learn the features based on different time resolutions. Each convolution layer has padding of $n$, which depend on kernel size, as given by the equation $n = (kernel\_size - 1)/2$. ReLU has been used as an activation function for each convolutional layer and can be defined as $ReLU(x) = max(0,x)$. The activation function in CNN has the two main advantages. First, the activation function convert the output to a scaled range which helps fast training of the network. Secondly, the combination of activation function help the network learn more complex non-linear patterns from input. Each convolutional branch has a dropout layer as a regularizer. During training, the dropout layers make some outputs from previous layer randomly ignored. This makes training process noisy and force the layers to co-adapt for mistakes made by prior layers which ultimately makes the network more robust.

The output features from the convolutional and residual branches are concatenated at the end which results in the output features of size $13 \times 12$ extracted by feature extraction module. The extracted features from each feature extraction module are further concatenated and then flattened in the consecutive layers. These flattened feature vector of size $780$ and the two parameters of environment temperature ($env_{temp}$) and battery's initial SOC ($batt_{soc}$) are concatenated to form the final feature vector of size $782$ which is then passed through two successive fully connected layers to get the final output of estimated power consumption of size $10$ for the 10 second interval.

\subsection{Re-sampler Module}\label{SubSec:Re-sampler Module}
The Re-sampler module up-sample or down-sample the input parameters to match the frequency of output generated by PCE Module. As discussed in the PCE module, the estimated power consumption output generated is of 1 Hz frequency i.e. one data point for each second, so it is necessary to re-sample the input parameters to match the frequency of 1 Hz before giving them as input to the Fine Tuner module. The input parameters namely, environment temperature and battery's initial SOC, as mentioned in PCE module are considered as constant for each 10 second interval. Due to this, the input parameters other than environment temperature and battery's initial SOC are down-sampled from 10 Hz to 1 Hz and the input parameters of environment temperature and battery's initial SOC are up-sampled to 1 Hz by repeating the same value 10 times.

\subsection{Fine Tuner Module}\label{SubSec:Fine Tuner Module}
The fine tuner module takes eight input parameters namely, vehicle's speed, road elevation, vehicle's acceleration, auxiliary loads, wind speed, environmental temperature, batter's initial SOC and estimated power consumption. First seven of these parameters are re-sampled by the Re-sampler module and the last one is estimated by the PCE module. The main purpose of fine tuner module is to fine tune the estimated power consumption based on the other input parameters by reducing the error in prediction. Multiple experiments were performed with different regression models and it has been found that Bagged Decision Trees (BDT) performed better than other regression models. The BDT is a collection of multiple decision trees ensembled together using bagging. Bagging or bootstrap aggregation is an ensemble learning technique proposed by Breiman \cite{Breiman1996}. The main objective of bagging is to reduce the variance of learners, in this case the decision trees. This is achieved by combining the output of multiple learners through averaging which ultimately help to achieve high prediction accuracy. In BDT, each learner is a separate decision tree and each decision tree is trained separately on subset of training dataset. For this, the training dataset is divided into multiple subsets using the bootstrap resampling. In bootstrap resampling, samples are selected from the training dataset with replacement. The output from all the separate decision trees are then averaged to obtain the final predictions which is in this case the fine tuned estimated power consumption. In this work, the number of trees and other parameters of the BDT are optimized by minimizing the cross validation error. The optimized model contains 10 weak tree learners, ensembled together in BDT with minimum prediction accuracy.

\subsection{State of Charge Calculator Module}\label{SubSec:StateofChargeCalculatorModule}
The fine tuned estimated power consumption obtained from Fine Tuner module is used by the State of Charge Calculator Module to calculate the remaining state of charge of battery. For this, the following equation, given in \cite{Kularatna2015}, can be used:

\begin{equation}
SOC_t = SOC_0 - \left( \frac{\int_0^t Est_{pow}dt}{E_{cap}} \times 100 \right)
\end{equation}

where $SOC_0$, $SOC_t$ represent the battery's initial state of charge and state of charge at time $t$. $Est_{pow}$ is the fine tuned estimated power consumption for the time interval and $E_{cap}$ represent the battery's rated energy capacity. In the current case as each of the small partitioned time series is of 10 sec duration, value of $t$ will be 10. So, the system requires the initial SOC only for the first partitioned time series and after that it can calculate the SOC based on the energy consumed by the vehicle and the calculated SOC can be used as initial SOC for next time series partition.

\section{Experimental Setup}\label{Sec:ExperimentalSetup}
In this section, the details of dataset, hyperparameters and other information related to the training of the proposed approach has been discussed.

\subsection{Datasets}\label{SubSec:DataSets}
For training, testing and validating the proposed approach data set from two different sources was used. The first source was Downloadable Dynamometer Database ($D^3$) \cite{D3ANL} which contains data obtained by performing a number of dynamometer tests at Argonne National Laboratory (ANL) of Advanced Powertrain Research Facility (APRF) on number of electric vehicles. These tests were conducted for number of drive cycles at 0\% road elevation under different environmental conditions.

As the data available at $D^3$ was very small and was not sufficient for training, testing and validating the approach, a large amount of dataset was obtained from a Nissan Leaf's simulated model provided in a simulation tool named FASTSim (Future Automotive Systems Technology Simulator) \cite{2015-01-0973}. The simulation model used Nissan Leaf's vehicle specific parameters which can be obtained from \cite{factNissan2013, burress2012benchmarking, MODI2019}. Dataset for other electric vehicles can also be obtained by developing similar simulated models for other electric vehicles subject to the availability of vehicle data from manufacturer like efficiency curve of motor, internal resistance of battery etc. Dataset was generated using the simulated model for 5 different temperatures (from $-5^\circ C$ to $35^\circ C$ with interval of $10^\circ C$), 10 road elevation profiles (road grade varies from -20\% to 20\%), 4 different initial state of charge of battery (from 30\% to 90\% with interval of 20\%), 40 drive cycles (like UDDS (Urban Dynamometer Driving Schedule) and SFTP (Supplemental Federal Test Procedures) etc) and 8 wind speed profiles. Wind speed has been categorized into 13 different categories according to Beaufort scale \cite{jebson2007fact}, namely, Calm (< 0.5 m/s), Light Air (0.5 - 1.5 m/s), Light Breeze (1.6 - 3.3 m/s),  Gentle Breeze (3.4 - 5.5 m/s),  Moderate Breeze (5.5 - 7.9 m/s),  Fresh Breeze (8 - 10.7 m/s),  Strong Breeze (10.8 - 13.8 m/s), Near Gale (13.9 - 17.1 m/s), Gale (17.2 - 20.7 m/s), Strong Gale (20.8 - 24.4 m/s), Storm (24.5 - 28.4 m/s), Violent Storm (28.5 - 32.6 m/s) and Hurricane ($\geq$ 32.7 m/s). Out of these 13 categories first 8 were used for data generation from simulated model because the remaining categories are not preferable conditions for driving. It is to be noted that more data can be generated by varying the environmental and road conditions and using new drive cycles.

From now onwards, the dataset obtained using the simulated model of FASTSim and the dataset downloaded from $D^3$ will be denoted as $DS-I$ and $DS-II$, respectively. Both the datasets have time series data which is recorded at frequency of 10 Hz i.e. for each second there are 10 readings. In order to train, validate and test the proposed approach datasets were partitioned into a number of partitions each of 10 seconds interval. So, in total the dataset $DS-I$ and $DS-II$ were partitioned into approximately 17 lacs and 3500 partitions, respectively. 70\% of the partitions of dataset $DS-I$ were used for training the proposed approach and the remaining partitions, i.e. 30\%, were used for validation. Henceforth, the training and validation dataset will be represented by $DS-I_{tr}$ and $DS-I_{val}$, respectively. The testing of the proposed approach was done using dataset $DS-II$.

\subsection{Data Preprocessing}\label{SubSec:DataPreprocessing}
As discussed in Section \ref{SubSec:DataSets}, both the datasets have data of multiple parameters recorded at 10 Hz frequency. For experimental purpose, the time series data from the datasets were divided into smaller partitions each of 10 sec interval i.e. each partition has 100 readings. Time series data for each parameter (say $z$) from the dataset was normalized, using the Eq. \eqref{Eq:Normalize}, into the range of [0,1] before using it for training, validation or testing the proposed approach.

\begin{equation}\label{Eq:Normalize}
    \hat{z^i} = \frac{z^i - \min(z)}{\max(z) - \min(z)}
\end{equation}

In above equation, $\hat{z^i}$, $z^i$, $\min(z)$ and $\max(z)$ represent the $i^{th}$ normalized partition of time series $z$, $i^{th}$ partition of time series $z$, minimum and maximum values of time series $z$, respectively.

\subsection{Training}\label{SubSec:Training}
In the proposed model there are mainly two modules, namely, PCE Module and Fine Tuner Module. The PCE module was implemented in Python using PyTorch APIs and the Fine Tuner module was implemented in MATLAB 2019a. PyTorch packages,  \emph{torch.nn} and \emph{torch.optim}, was used to define the multi-channel CNN architecture for PCE module and loss functions used for learning. It also provides a number of APIs for training and testing the created model. For training the BDT of Fine Tuner Module an addon, named Regression Learner App, provided in MATLAB was used. It can be used for training number of different regression models in MATLAB. The implementation code for the proposed approach along with sample data has been provided on GitHub (\emph{https://github.com/shatrughanmodi/CNN-BDT}). After training the PCE and Fine Tuner modules separately, both the modules were integrated by calling MATLAB code from Python. The training of modules was done on a system with 8 GB RAM and Intel i5 Processor. While training the modules, the feedback loop, in which the remaining SOC calculated after each 10 sec interval is fed to the initial SOC of next interval, was not used, but the feedback loop was used while testing the proposed approach.

First, the PCE module was trained using the $70\%$ of time series data from the dataset $DS-I$. For training the PCE module, the weights of all the convolutional and fully connected layers were initialized using the Xavier initialization \cite{glorot2010understanding}. It initializes the layer's weight from a random uniform distribution with limits of $\left[-\sqrt{\frac{6}{fan_{in}+fan_{out}}}, \sqrt{\frac{6}{fan_{in}+fan_{out}}}\right]$, where $fan_{in}$ and $fan_{out}$ are the number of input connections to the layer and number of output connections from the layer, respectively. The PCE module was trained with batch size of 64 for 3000 epochs using Adam optimization algorithm \cite{kingma2014adam}. The Adam optimizer has the combined advantages of two stochastic gradient descent algorithms i.e. AdaGrad and RMSProp. The Adam Optimizer uses the delta learning rule to update the weights of the network. It basically computes an exponential running average of the gradients and square of gradients. The decay rate of these running averages are controlled by the parameters $\beta_1$ and $\beta_2$ which were initialized to 0.9 and 0.999, respectively. The initial learning rate $\alpha$ was set to 0.001 and to avoid division by zero during training epsilon $\epsilon$ was set to $10^{-8}$. The Mean Square Error (MSE) was used as the loss function to minimize the MSE between the actual and estimated power consumption. To avoid overfitting, dropout layers were used as regularizers in each convolutional branch of feature extraction module. Number of experiments have been performed by varying the drop rate $p$ and it has been found that increasing the drop rate after certain threshold (in this case 0.2) does not reduce the testing error. After training the PCE module, the Fine Tuner Module was trained using the output obtained from PCE module and re-sampled input data obtained from the Re-sampler module.

The bagged decision tree of Fine Tuner Module was developed by ensembling the multiple decision trees. The number of trees and other parameters were optimized by using the bayesian optimization which tries to minimize the cross validation error. The model, optimized with bayesian optimization, has 10 weak tree learners which were ensembled to form a bagged decision tree with minimum prediction error.

\section{Results and Discussion}\label{Sec:Results}
\begin{figure}[h!]
  \centering
  \begin{subfigure}[]{0.32\textwidth}
    \centering
    \includegraphics[width=\textwidth]{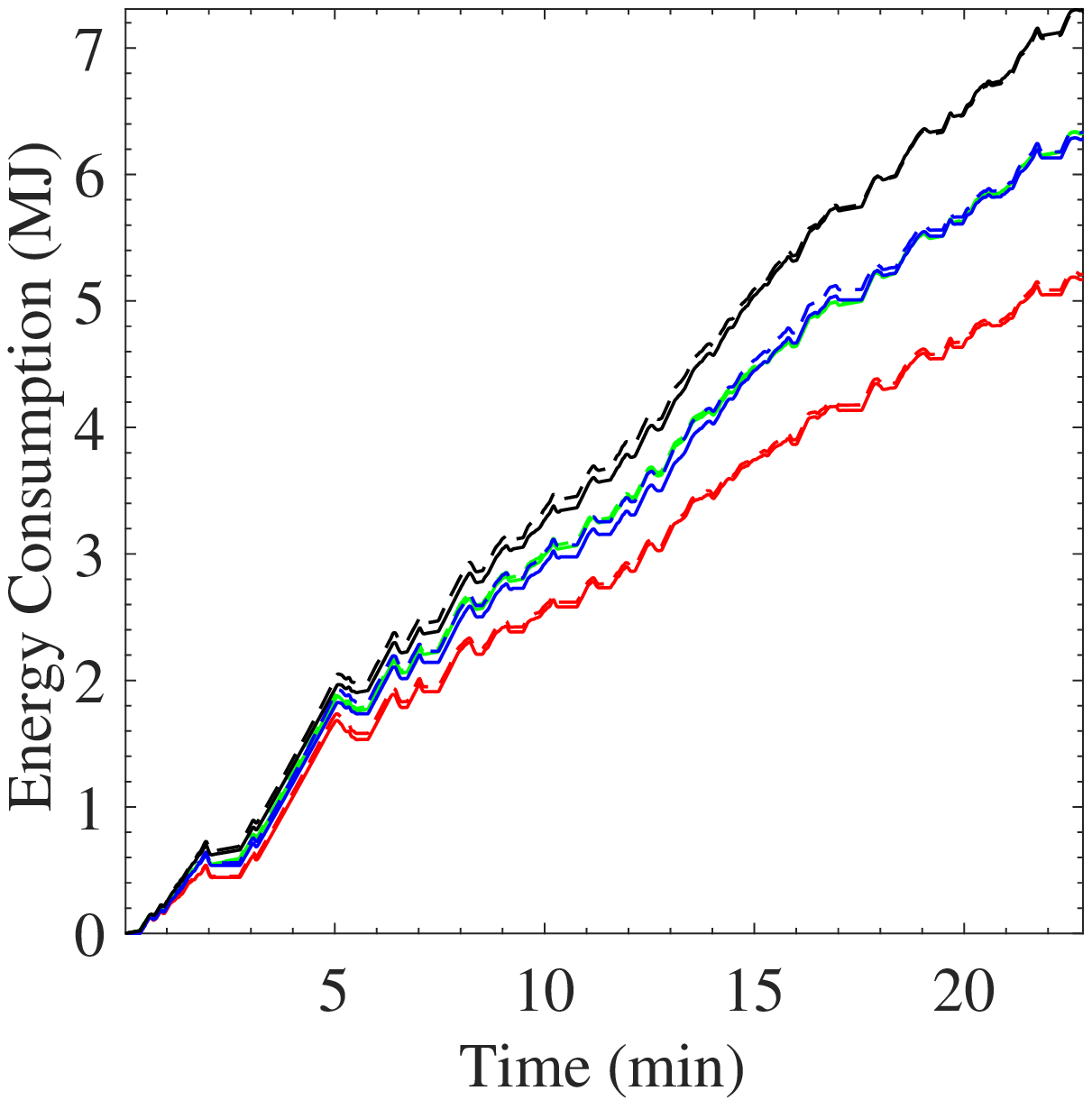}
    \caption{Results with grade profile 1 and temperature $-5^o C$}\label{SubFig:G0_T-5_C30}
  \end{subfigure}
  \hfill
  \begin{subfigure}[]{0.32\textwidth}
    \centering
    \includegraphics[width=\linewidth]{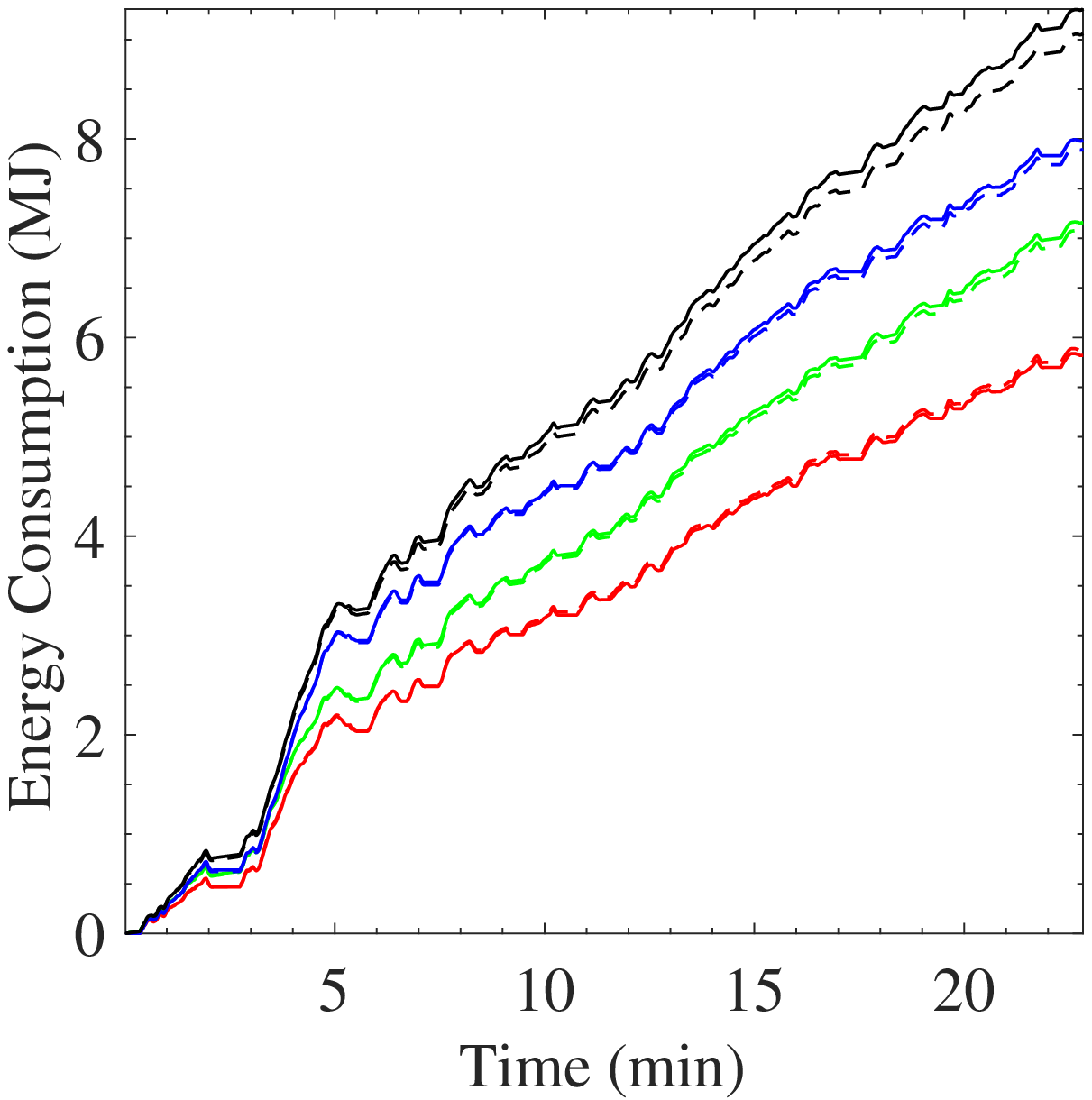}
    \caption{Results with grade profile 1 and temperature $15^o C$}\label{SubFig:G0_T15_C30}
  \end{subfigure}
  \hfill
  \begin{subfigure}[]{0.32\textwidth}
    \centering
    \includegraphics[width=\linewidth]{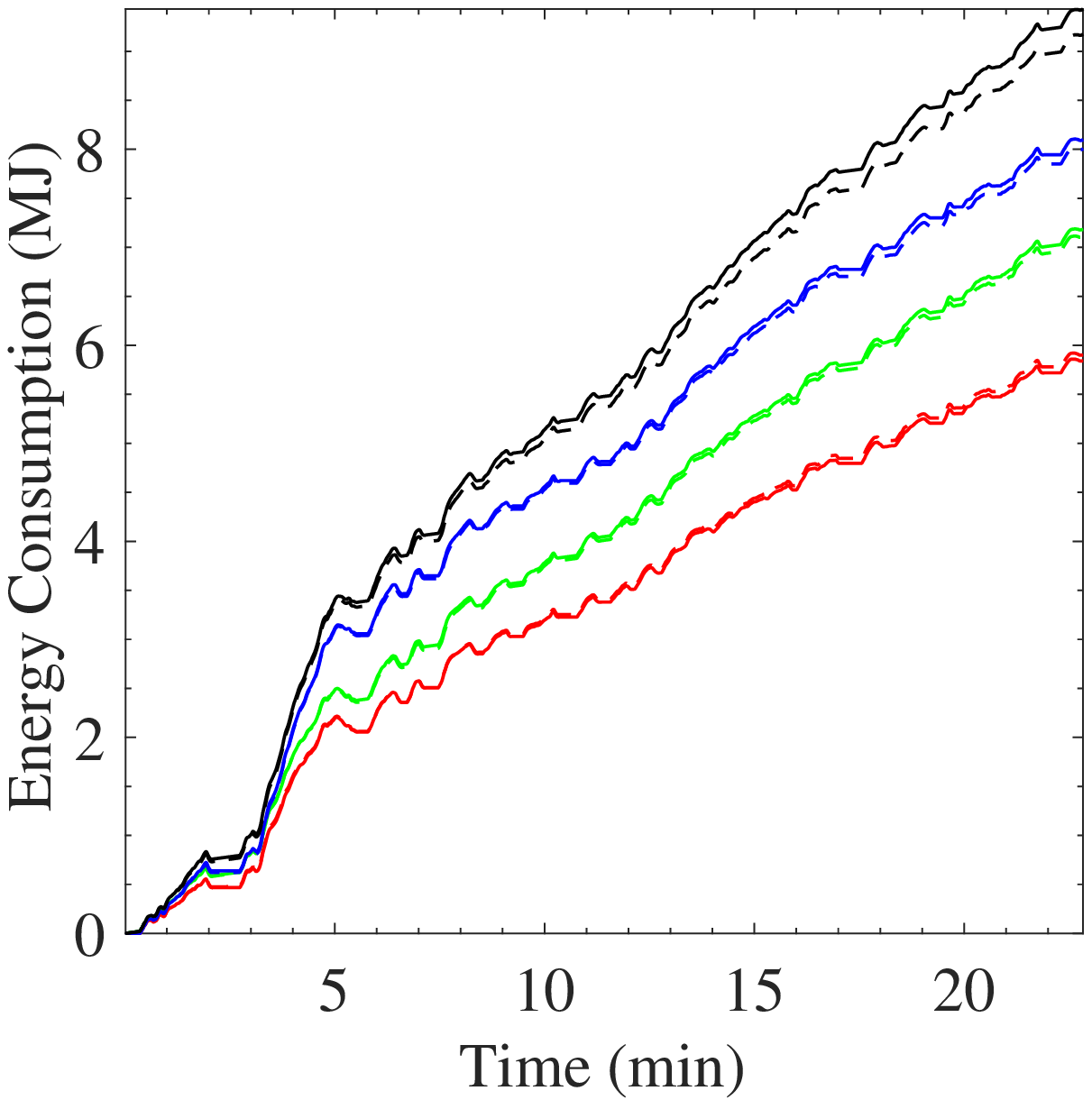}
    \caption{Results with grade profile 1 and temperature $35^o C$}\label{SubFig:G0_T35_C30}
  \end{subfigure}
  \begin{subfigure}[]{0.32\textwidth}
    \centering
    \includegraphics[width=\linewidth]{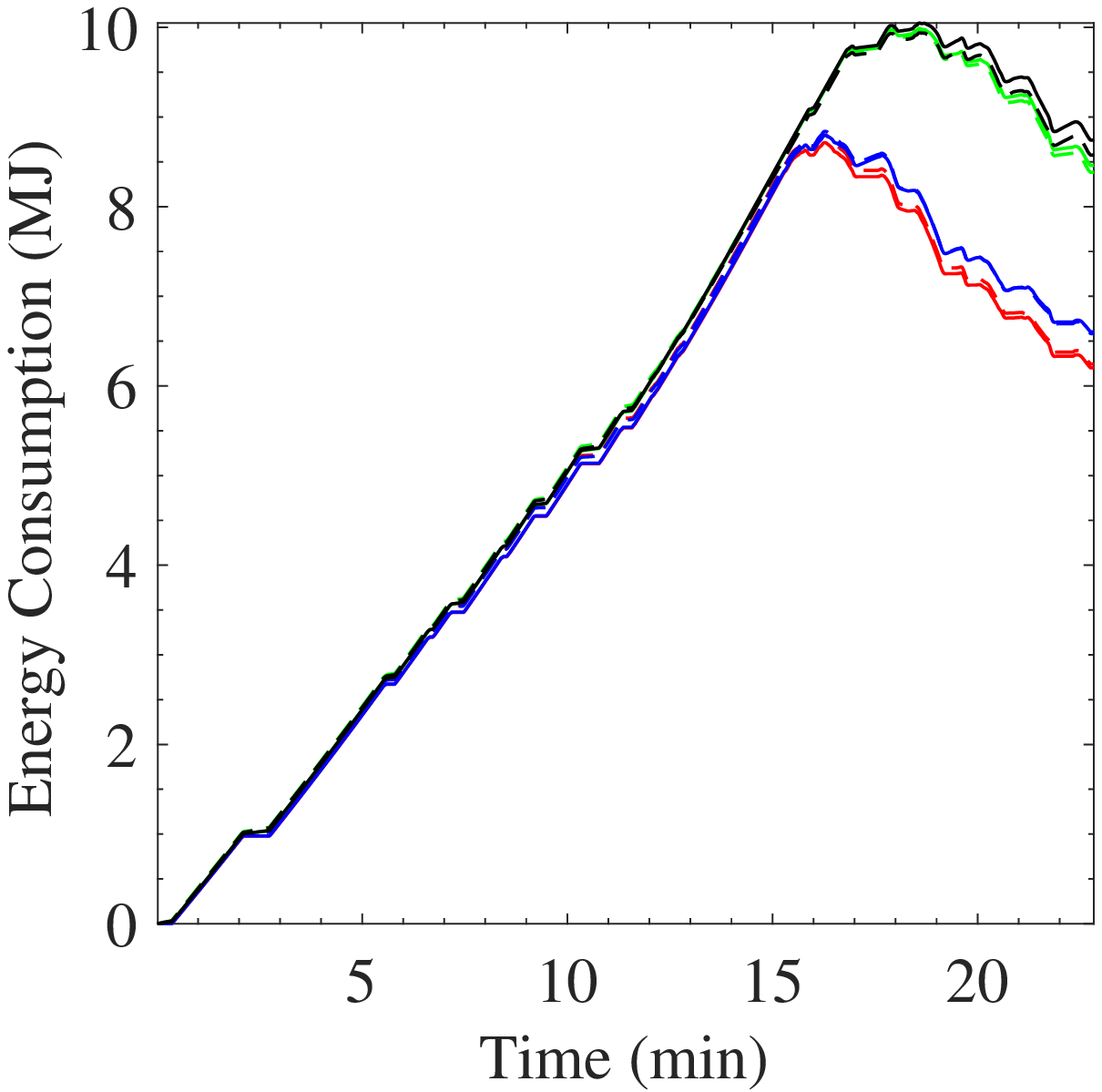}
    \caption{Results with grade profile 2 and temperature $-5^o C$}\label{SubFig:G10_T-5_C30}
  \end{subfigure}
  \hfill
  \begin{subfigure}[]{0.32\textwidth}
    \centering
    \includegraphics[width=\linewidth]{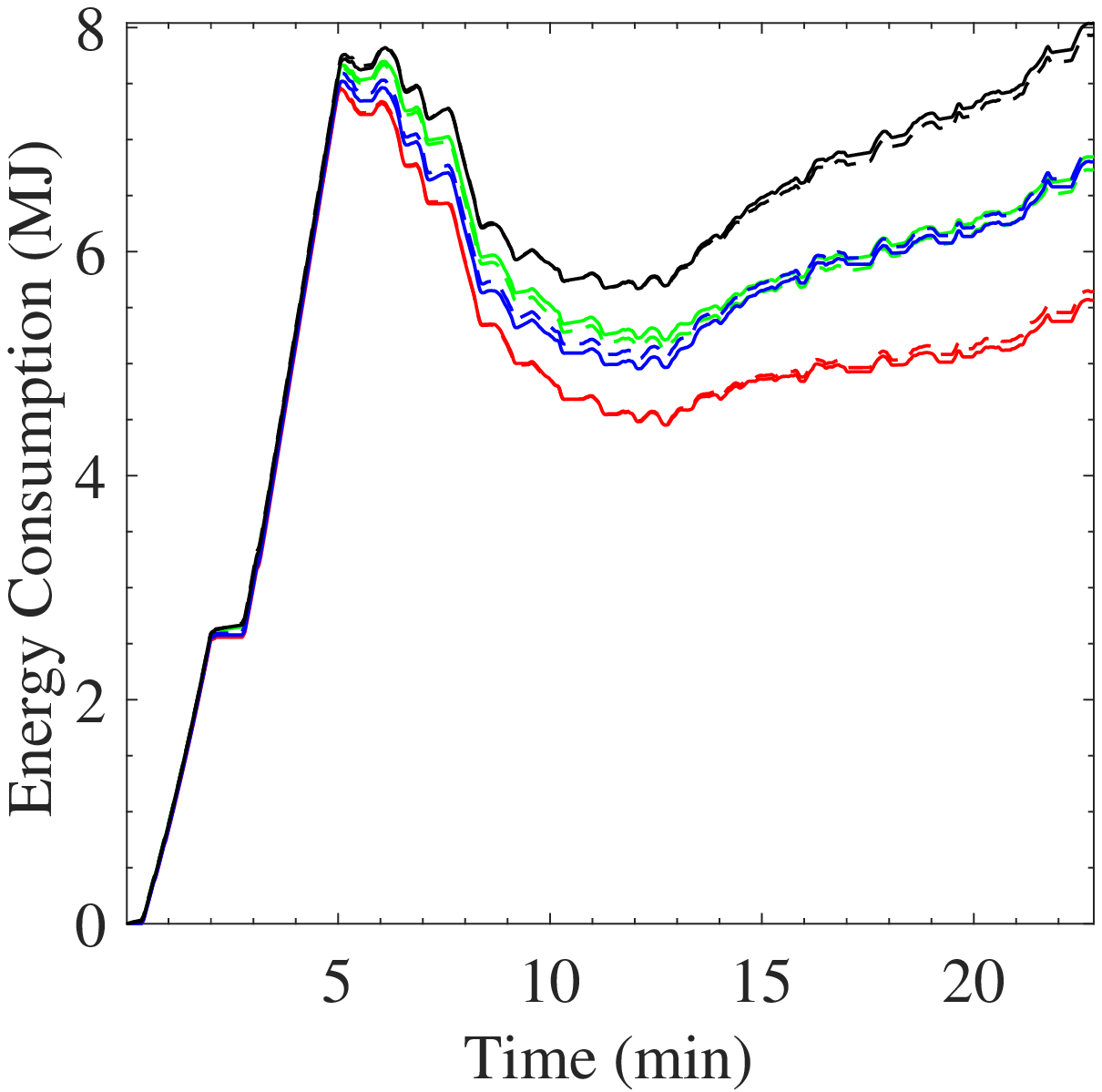}
    \caption{Results with grade profile 2 and temperature $15^o C$}\label{SubFig:G10_T15_C30}
  \end{subfigure}
  \hfill
  \begin{subfigure}[]{0.32\textwidth}
    \centering
    \includegraphics[width=\linewidth]{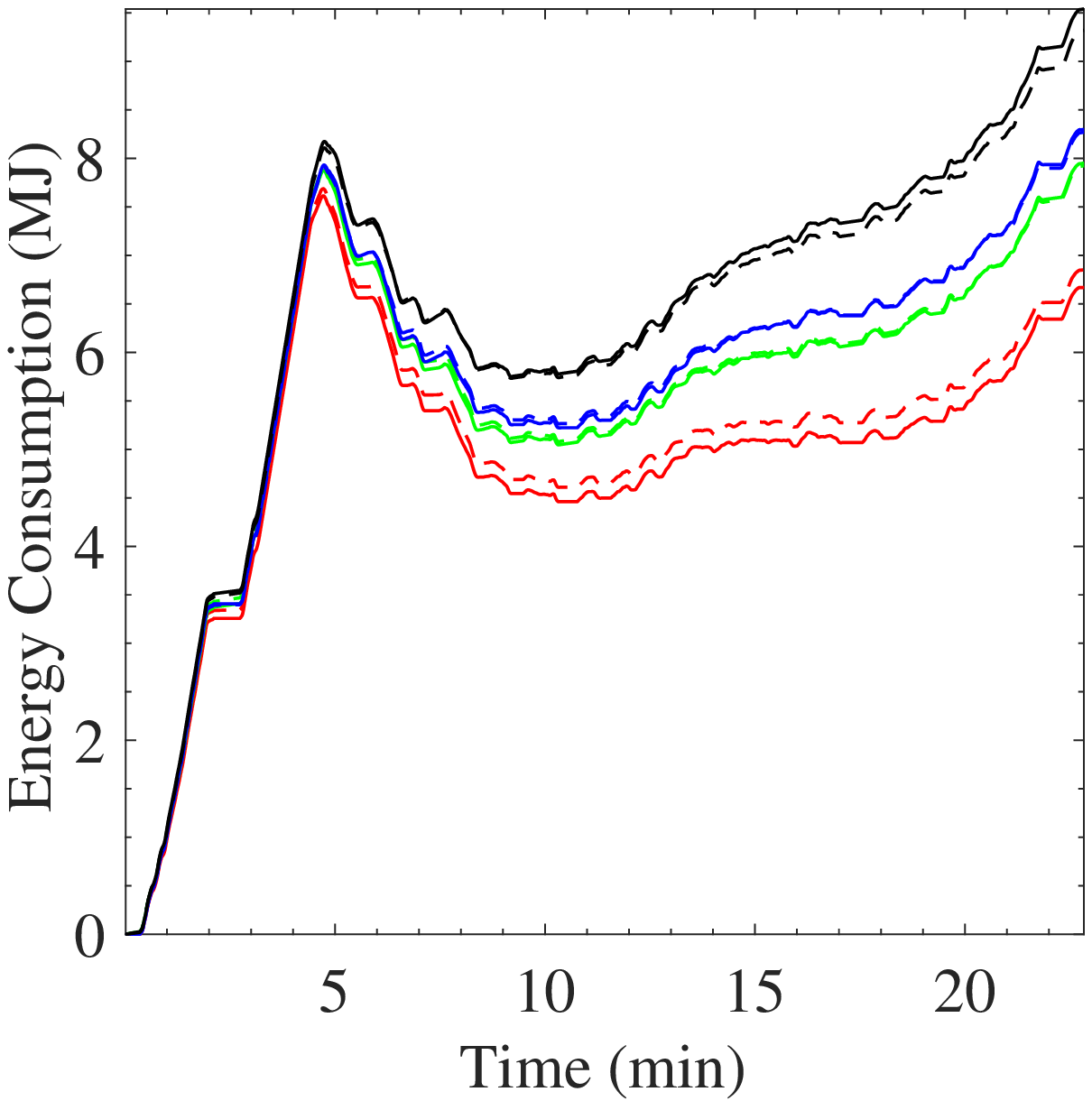}
    \caption{Results with grade profile 2 and temperature $35^o C$}\label{SubFig:G10_T35_C30}
  \end{subfigure}
  \begin{subfigure}[]{0.32\textwidth}
    \centering
    \includegraphics[width=\linewidth]{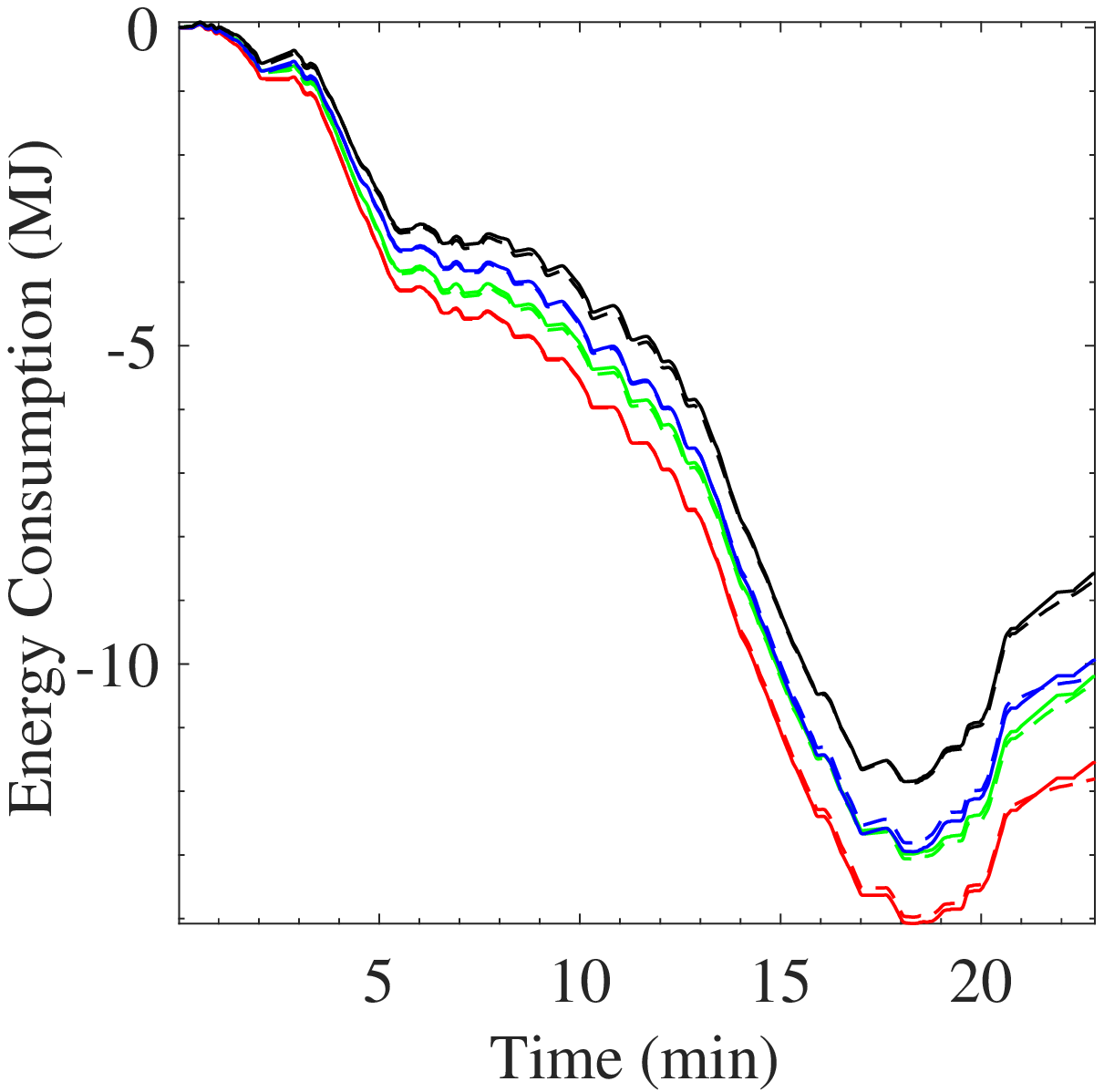}
    \caption{Results with grade profile 3 and temperature $-5^o C$}\label{SubFig:G25_T-5_C30}
  \end{subfigure}
  \hfill
  \begin{subfigure}[]{0.32\textwidth}
    \centering
    \includegraphics[width=\linewidth]{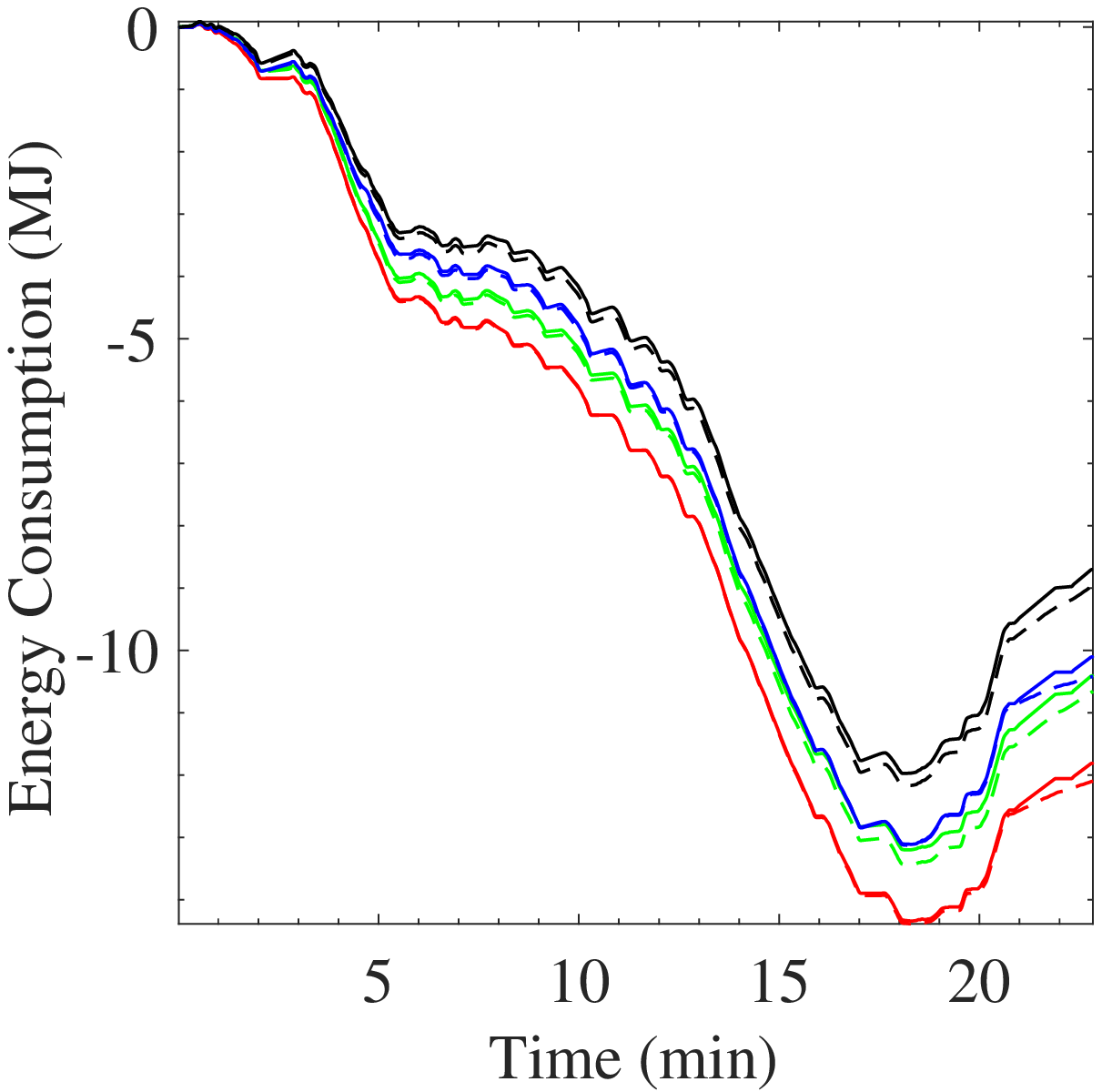}
    \caption{Results with grade profile 3 and temperature $15^o C$}\label{SubFig:G25_T15_C30}
  \end{subfigure}
  \hfill
  \begin{subfigure}[]{0.32\textwidth}
    \centering
    \includegraphics[width=\linewidth]{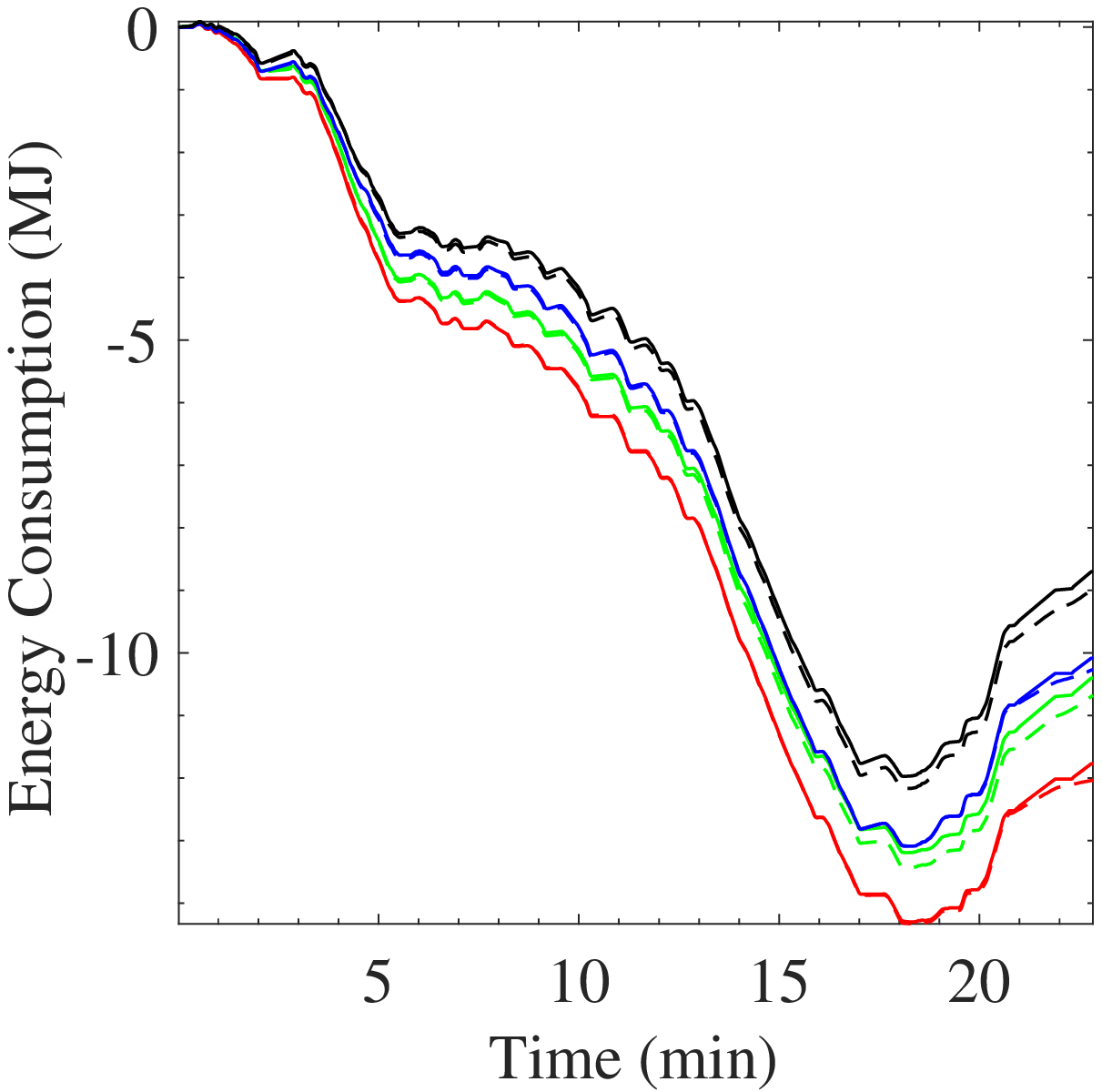}
    \caption{Results with grade profile 3 and temperature $35^o C$}\label{SubFig:G25_T35_C30}
  \end{subfigure}

  \caption{Energy consumption estimation for UDDS drive cycle at initial SOC of 30\% under different conditions. Legend: Actual (Solid Line) / Predicted (Dashed Line) energy consumption under: air speed profile 1 with auxiliary load profile 1 (\protect\solidcircle[red]) / auxiliary load profile 2 (\protect\solidcircle[green]), air speed profile 2 with auxiliary load profile 1 (\protect\solidcircle[blue]) / auxiliary load profile 2 (\protect\solidcircle[black]).}\label{Fig:EnergyConsumptionComparisonAtSOC30}
\end{figure}

\begin{figure}[h!]
  \centering
  \begin{subfigure}[]{0.32\textwidth}
    \centering
    \includegraphics[width=\textwidth]{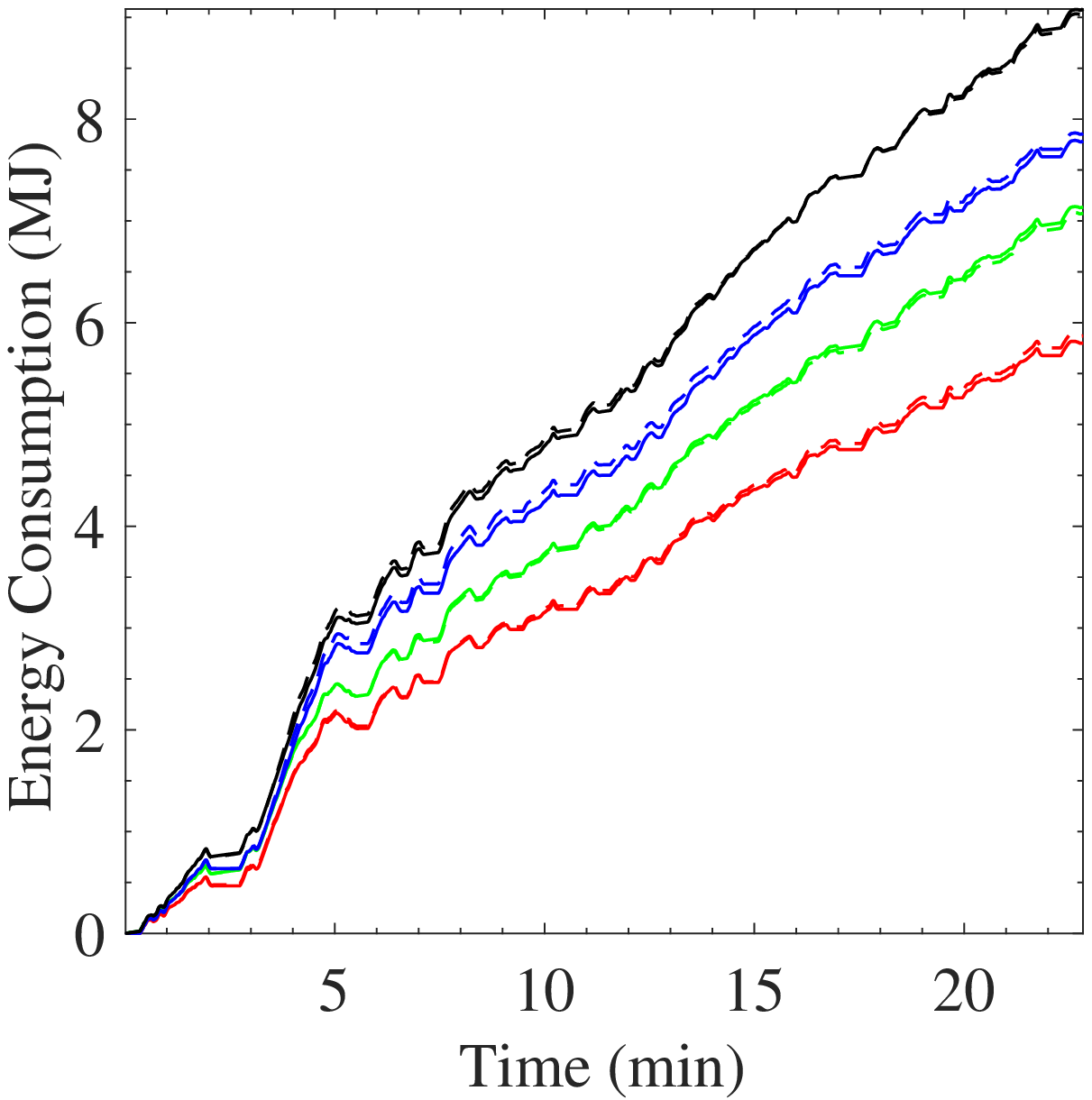}
    \caption{Results with grade profile 1 and temperature $-5^o C$}\label{SubFig:G0_T-5_C70}
  \end{subfigure}
  \hfill
  \begin{subfigure}[]{0.32\textwidth}
    \centering
    \includegraphics[width=\linewidth]{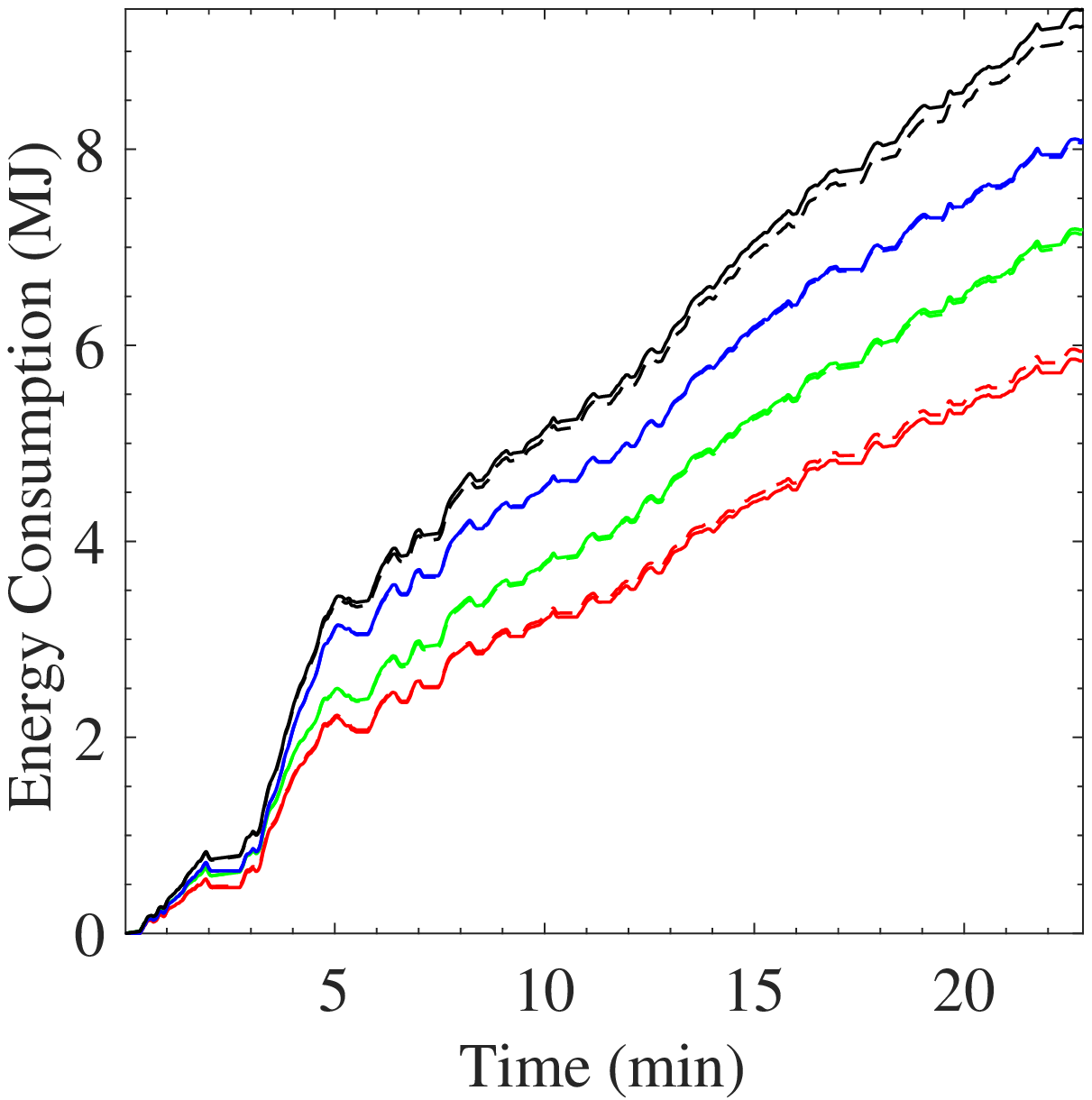}
    \caption{Results with grade profile 1 and temperature $15^o C$}\label{SubFig:G0_T15_C70}
  \end{subfigure}
  \hfill
  \begin{subfigure}[]{0.32\textwidth}
    \centering
    \includegraphics[width=\linewidth]{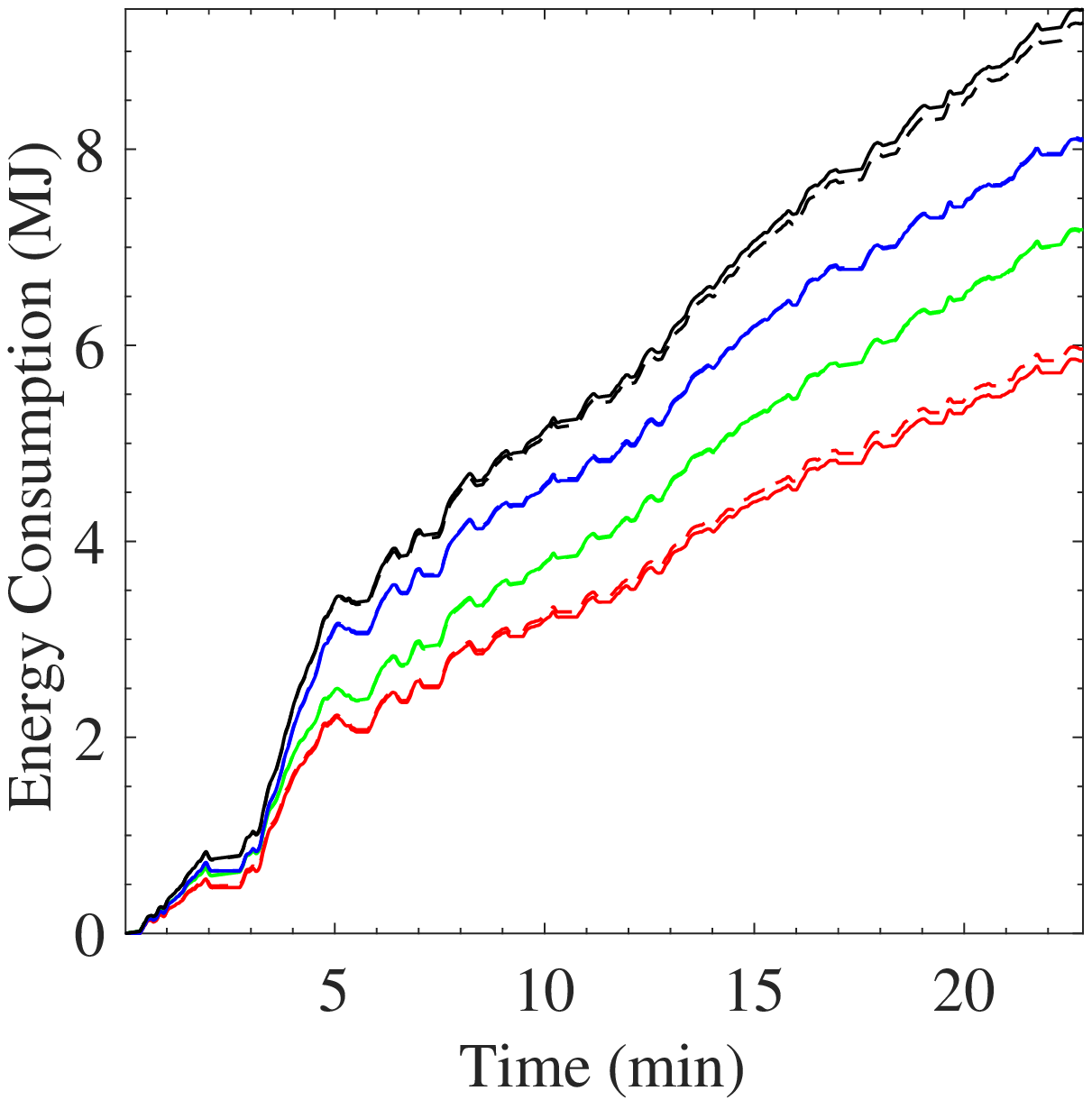}
    \caption{Results with grade profile 1 and temperature $35^o C$}\label{SubFig:G0_T35_C70}
  \end{subfigure}
  \begin{subfigure}[]{0.32\textwidth}
    \centering
    \includegraphics[width=\linewidth]{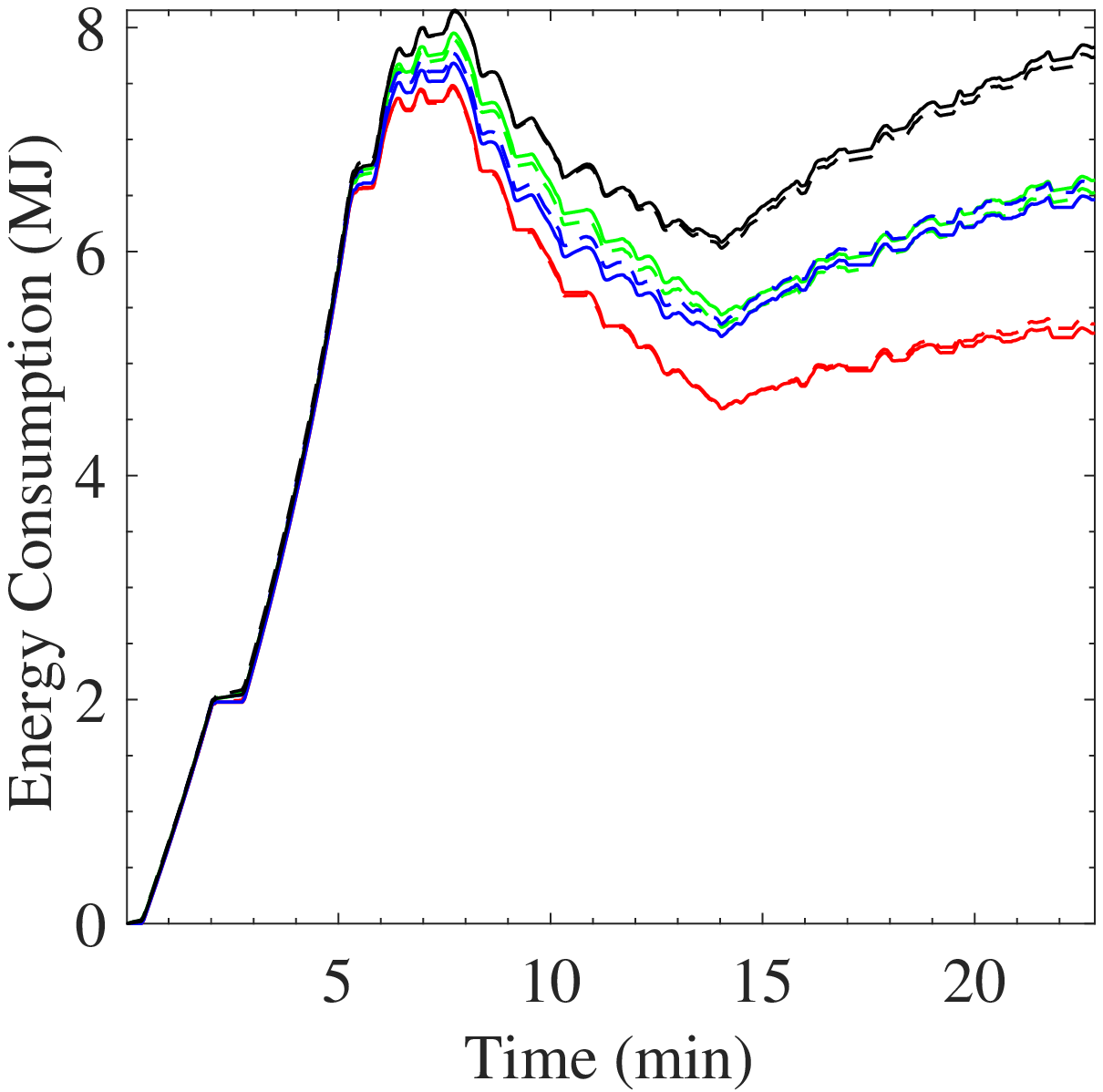}
    \caption{Results with grade profile 2 and temperature $-5^o C$}\label{SubFig:G10_T-5_C70}
  \end{subfigure}
  \hfill
  \begin{subfigure}[]{0.32\textwidth}
    \centering
    \includegraphics[width=\linewidth]{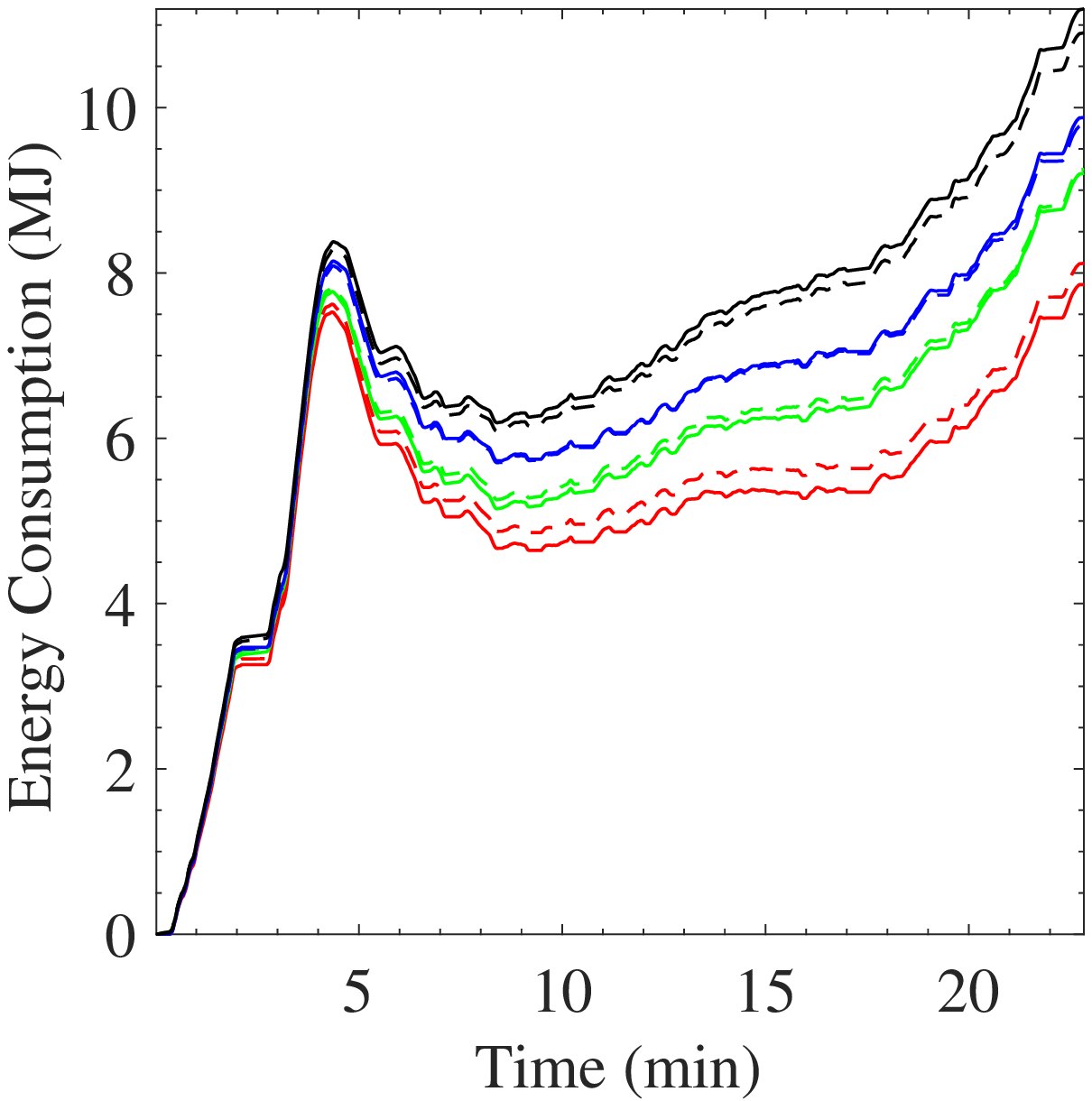}
    \caption{Results with grade profile 2 and temperature $15^o C$}\label{SubFig:G10_T15_C70}
  \end{subfigure}
  \hfill
  \begin{subfigure}[]{0.32\textwidth}
    \centering
    \includegraphics[width=\linewidth]{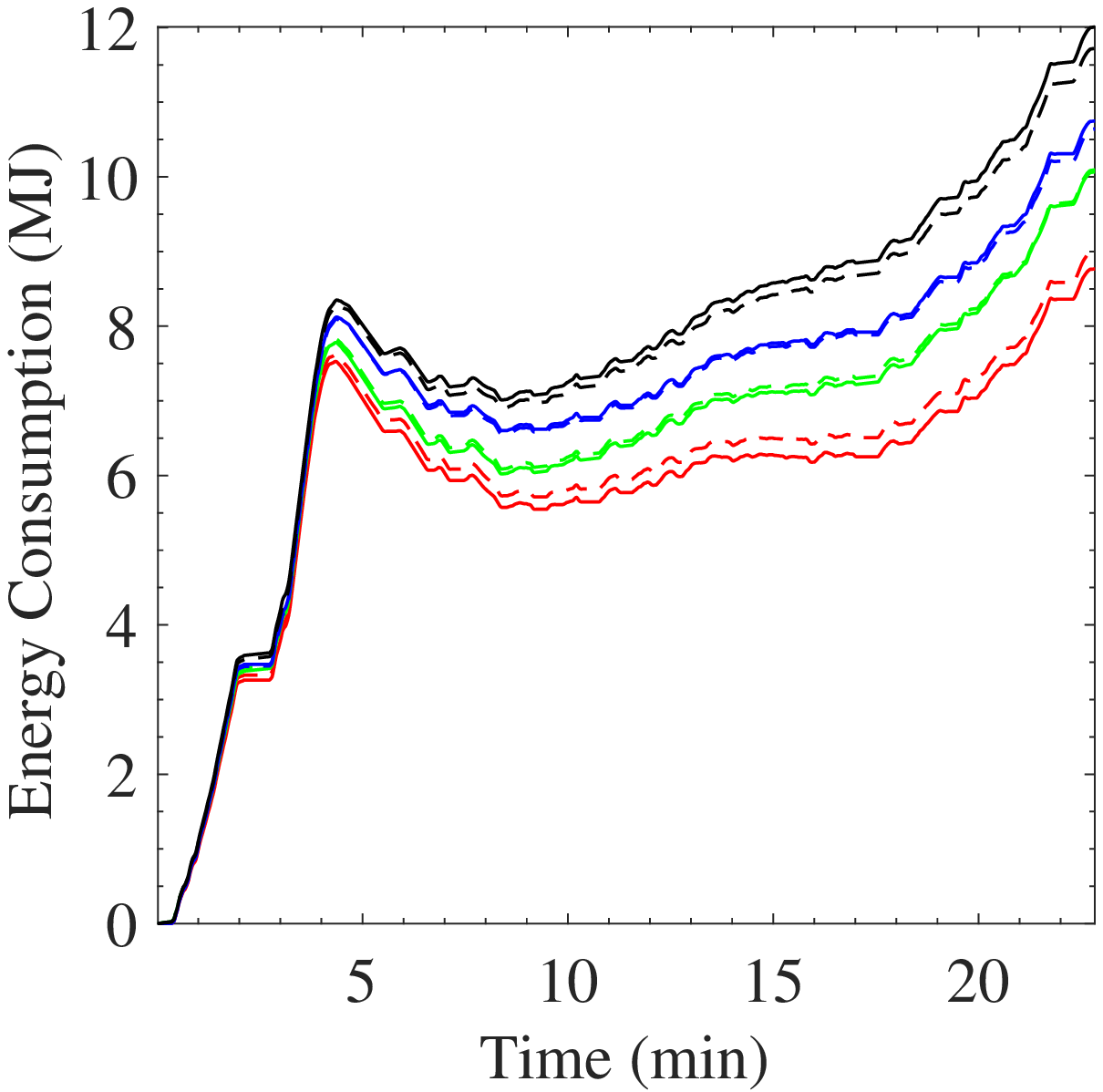}
    \caption{Results with grade profile 2 and temperature $35^o C$}\label{SubFig:G10_T35_C70}
  \end{subfigure}
  \begin{subfigure}[]{0.32\textwidth}
    \centering
    \includegraphics[width=\linewidth]{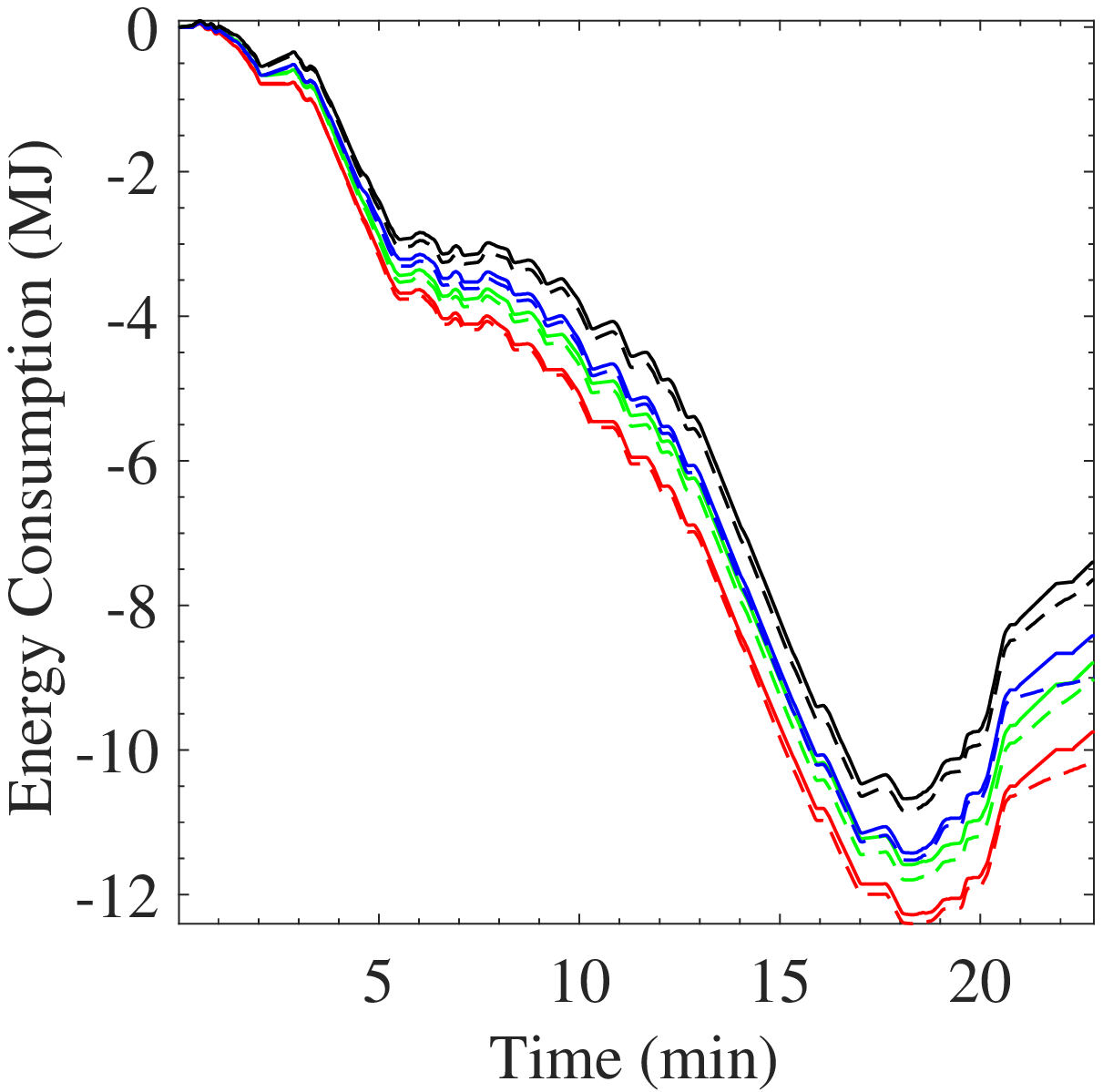}
    \caption{Results with grade profile 3 and temperature $-5^o C$}\label{SubFig:G25_T-5_C70}
  \end{subfigure}
  \hfill
  \begin{subfigure}[]{0.32\textwidth}
    \centering
    \includegraphics[width=\linewidth]{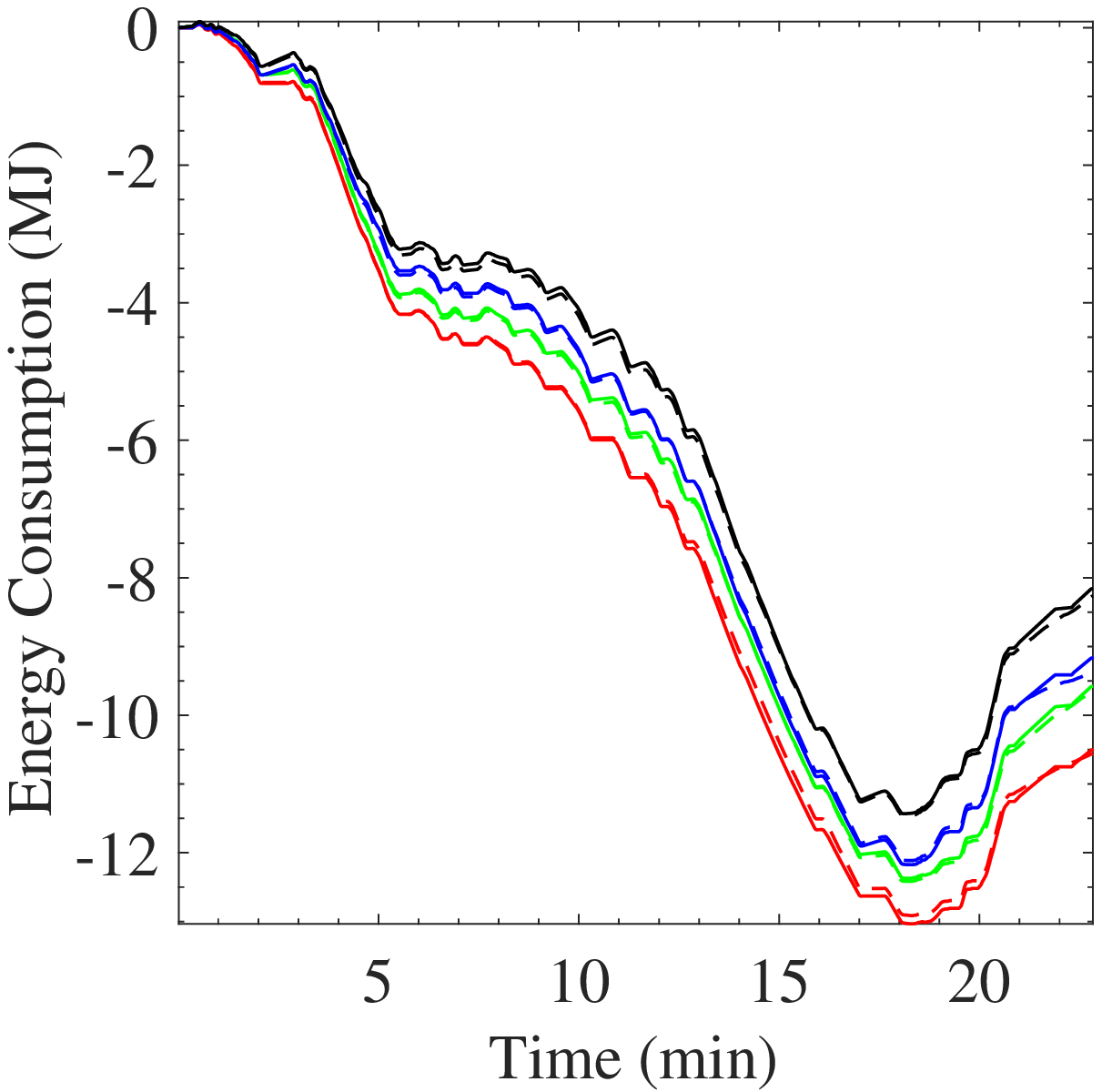}
    \caption{Results with grade profile 3 and temperature $15^o C$}\label{SubFig:G25_T15_C70}
  \end{subfigure}
  \hfill
  \begin{subfigure}[]{0.32\textwidth}
    \centering
    \includegraphics[width=\linewidth]{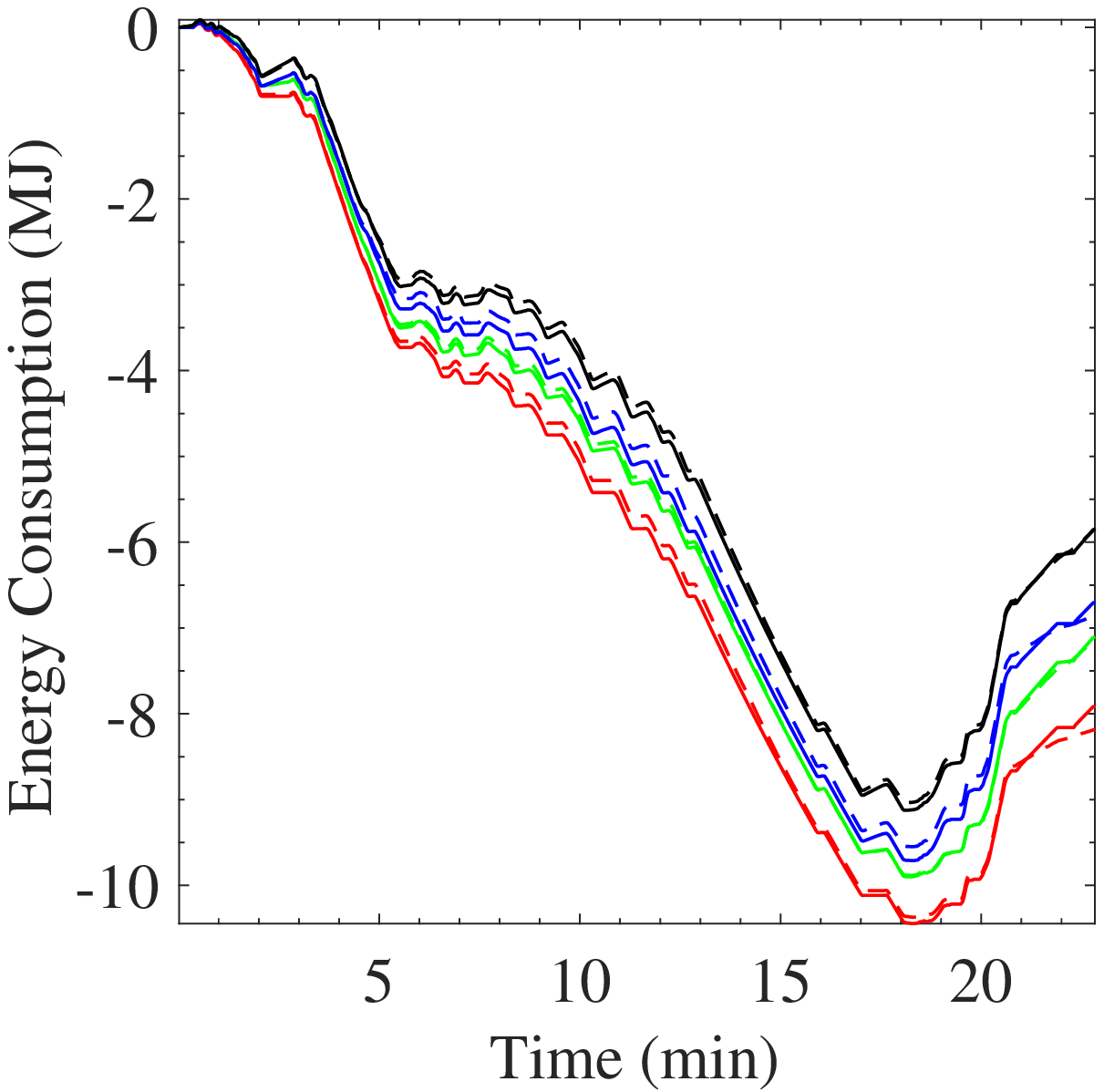}
    \caption{Results with grade profile 3 and temperature $35^o C$}\label{SubFig:G25_T35_C70}
  \end{subfigure}

  \caption{Energy consumption estimation for UDDS drive cycle at initial SOC of 70\% under different conditions. Legend: Actual (Solid Line) / Predicted (Dashed Line) energy consumption under: air speed profile 1 with auxiliary load profile 1 (\protect\solidcircle[red]) / auxiliary load profile 2 (\protect\solidcircle[green]), air speed profile 2 with auxiliary load profile 1 (\protect\solidcircle[blue]) / auxiliary load profile 2 (\protect\solidcircle[black]).} \label{Fig:EnergyConsumptionComparisonAtSOC70}
\end{figure}

In this section, the testing results obtained using proposed architecture have been discussed. Figure \ref{Fig:EnergyConsumptionComparisonAtSOC30} and \ref{Fig:EnergyConsumptionComparisonAtSOC70}, show energy consumption prediction comparison of the proposed approach with the actual energy consumption of EV under different conditions for UDDS (Urban Dynamometer Driving Schedule) drive cycle with initial battery's SOC level at 30\% and 70\%, respectively. The different profiles of road grade, air speed, auxiliary load and vehicle speed, which were used to obtain the results shown in above mentioned figures, have been given in Table \ref{Tab:ParameterProfiles}.

\begin{table}[h!]
    \centering
    
    \caption{Different Parameter Profiles}
    \label{Tab:ParameterProfiles}
    
    \begin{tabular}{|c|c|c|c|c|c|}
        \hline
        Parameter Name                       & Profile  & Minimum   & Maximum   & Mean      & Standard Deviation \\ \hline
        Vehicle Speed (m/s)                  & UDDS     & 0         & 25.347    & 8.747     & 6.576              \\ \hline
        \multirow{2}{*}{Auxiliary Load (W)}  & Profile 1& 0         & 0         & 0         & 0                  \\ \cline{2-6}
                                             & Profile 2& 952.941   & 1124.705  & 977.649   & 41.605             \\ \hline
        \multirow{2}{*}{Air Speed (m/s)}     & Profile 1& 0.084     & 0.214     & 0.150     & 0.019              \\ \cline{2-6}
                                             & Profile 2& 11.305    & 13.187    & 12.299    & 0.353              \\ \hline
        \multirow{3}{*}{Road Grade (\%)}     & Profile 1& 0         & 0         & 0         & 0                  \\ \cline{2-6}
                                             & Profile 2& -18.765   & 17.696    & 1.198     & 9.107              \\ \cline{2-6}
                                             & Profile 3& -17.953   & 10.262    & -8.064    & 9.883              \\ \hline
    \end{tabular}
    
\end{table}

There are number of observations that can be drawn from the Figures \ref{Fig:EnergyConsumptionComparisonAtSOC30} and \ref{Fig:EnergyConsumptionComparisonAtSOC70}. From a single sub-figure of these figures, it can be observed that how auxiliary loads and air speed influence the energy consumption of EV. For instance, observe Figure \ref{SubFig:G0_T35_C30}, energy consumption for four combinations of air speed and auxiliary loads have been presented each with different color. It can be seen that when the wind flows at higher speed as in air speed profile 2 and also auxiliary loads have been applied as in auxiliary load profile 2 the energy consumption is highest as compared to the energy consumption with other combination of air speed and auxiliary loads. There is approximately 50\% increase in energy consumption under air speed profile 2 and auxiliary load profile 2 as compared to under air speed profile 1 and auxiliary load profile 1 (i.e. no auxiliary load). So, a significant amount of energy gets consumed to overcome the aerodynamic drag and fulfill the demand of auxiliary loads.

The effect of road grade profile can be seen clearly by comparing the energy consumption within a column of the Figures \ref{Fig:EnergyConsumptionComparisonAtSOC30} and \ref{Fig:EnergyConsumptionComparisonAtSOC70}, where all other parameters are same but only road grade varies. For instance, consider the middle column of Figure \ref{Fig:EnergyConsumptionComparisonAtSOC70}, the energy consumption for air speed profile 2 and auxiliary load profile 2 after completing the drive cycle is about 9 MJ, 11 MJ and -8 MJ in Figure \ref{SubFig:G0_T15_C70}, \ref{SubFig:G10_T15_C70} and \ref{SubFig:G25_T15_C70}, respectively. Energy consumption for grade profile 2 is more as compared to grade profile 1 because grade profile 1 has zero mean and zero standard deviation i.e. no elevation/de-elevation whereas grade profile 2 has +ve mean i.e. mostly elevation. Similarly, energy consumption for grade profile 3 is -ve because mean grade for grade profile 3 is -ve i.e. de-elevation, so the energy will be generated which will be used to charge the battery, hence -ve energy consumption.

Similar to road grade, the effect of temperature can be observed by considering one of the rows of the Figures \ref{Fig:EnergyConsumptionComparisonAtSOC30} and \ref{Fig:EnergyConsumptionComparisonAtSOC70}, where only environment temperature is different for each sub-figure in a row and all other parameters are same. For instance, consider the second row of Figure \ref{Fig:EnergyConsumptionComparisonAtSOC30}, the difference in energy consumption due to change in temperature is clearly visible. There is a peak at approximately 5 min in Figures \ref{SubFig:G10_T15_C30} and \ref{SubFig:G10_T35_C30} whereas no such peak exist in Figure \ref{SubFig:G10_T-5_C30}. The main reason for this is that the battery's performance degrades with decrease in temperature as battery's capacity decreases at low temperature due to increase in internal resistance of battery. Due to this, battery can not provide enough power to the vehicle to reach the desired speed or acceleration and hence driver is forced to run the vehicle at low speed. The peak in the figures is due to high speed/acceleration demand which was successfully fulfilled at temperature of $15^\circ C$ and $35^\circ C$ but at environmental temperature of $-5^\circ C$ the battery was not able to provide the enough power. This behaviour is normally seen when the battery's SOC is low like in this case where initial SOC is 30\% but when the initial SOC is 70\%, as in second row of Figure \ref{Fig:EnergyConsumptionComparisonAtSOC70}, the energy consumption pattern at low temperature is quite similar to the energy consumption pattern at high temperature.

From the above discussion, it can be concluded that all of these parameters namely, vehicle speed, vehicle acceleration, road elevation, wind speed, auxiliary loads, environmental temperature and initial battery's SOC, are influencing the energy consumption of EV significantly. Also from the Figures \ref{Fig:EnergyConsumptionComparisonAtSOC30} and \ref{Fig:EnergyConsumptionComparisonAtSOC70}, it can be observed that the proposed approach successfully represent the non-linear influence of these parameters on energy consumption of EV with small deviation and can be used to estimate the energy consumption of EV in real-time.

\subsection{Cross Validation}\label{SubSec:CrossValidation}
Cross validation of the proposed approach has been performed to validate the generalizability and robustness of proposed methodology. There are several cross validation techniques and repeated $k$-fold cross validation is one of them, which is often used by the researchers to validate their approach. In the current work, repeated 10-fold cross validation has been performed. The performance of the proposed approach has been measured using the following metrics:

\begin{enumerate}[i)]
    \item \emph{Root Mean Square Error ($RMSE$):} It is a widely accepted and standardized metric to measure performance of a system by calculating the error rate. Following equation was used to calculate the $RMSE$ for the proposed approach:

        \begin{equation}\label{Eq:RMSE}
            RMSE = \sqrt{\frac{\sum^n_{i=1}(Act_{pow}^i - Est_{pow}^i)^2}{n}}
        \end{equation}

        where $Act_{pow}$ represent the actual power consumption, $Est_{pow}$ represent the power consumption estimated by the proposed approach and $n$ represent the total number of observations under consideration.

    \item \emph{Mean Absolute Error ($MAE$):} It is another standardized metric which provides an insight about the absolute deviation of estimation as compared to the actual value. Equation given below was used to measure the MAE between the estimated and actual power consumption:

        \begin{equation}\label{Eq:MAE}
            MAE = \frac{\sum^n_{i=1}|(Act_{pow}^i - Est_{pow}^i)|}{n}
        \end{equation}

        In above equation, same notations to represent the different variables of total number of used observations, actual and estimated power consumption has been used, as in Eq. \eqref{Eq:RMSE}.

    \item \emph{Correlation (Corr):} To measure the relationship between two variables statistically, correlation has been widely used. Value of correlation lie between [-1,1] for two variables. The negative value of correlation for two variables, say $x$ and $y$, indicates that both the variables show opposite behaviour to each other i.e. if $x$ increases then $y$ decreases and if $x$ decreases then $y$ increases. Zero correlation means the two variables have no relation among them and positive correlation between two variable $x$ and $y$ indicates the linear relation between the variables i.e. both the variables show similar behaviour which means $y$ increases if $x$ increases and $y$ decreases if $x$ decreases. So, for any system the correlation of estimated and actual variable must be close to 1 to be categorized as good. Following equation was used to calculate the correlation of estimated and actual power consumption:

        \begin{equation}\label{Eq:Correlation}
            Corr = \frac{\sum^n_{i=1}(Act_{pow}^i - \overline{Act_{pow}}) (Est_{pow}^i - \overline{Est_{pow}})} {\sqrt{\sum^n_{i=1}(Act_{pow}^i - \overline{Act_{pow}})^2 \sum^n_{i=1}(Est_{pow}^i - \overline{Est_{pow}})^2}}
        \end{equation}

        where $\overline{Act_{pow}}$ is the average actual power consumption and $\overline{Est_{pow}}$  is average estimated power consumption and other symbols represent the same variables as in Eq. \eqref{Eq:RMSE}.
\end{enumerate}

\begin{figure*}[h!]
  \centering
  
  \begin{subfigure}[t]{0.32\linewidth}
    \centering
    \includegraphics[width=\linewidth]{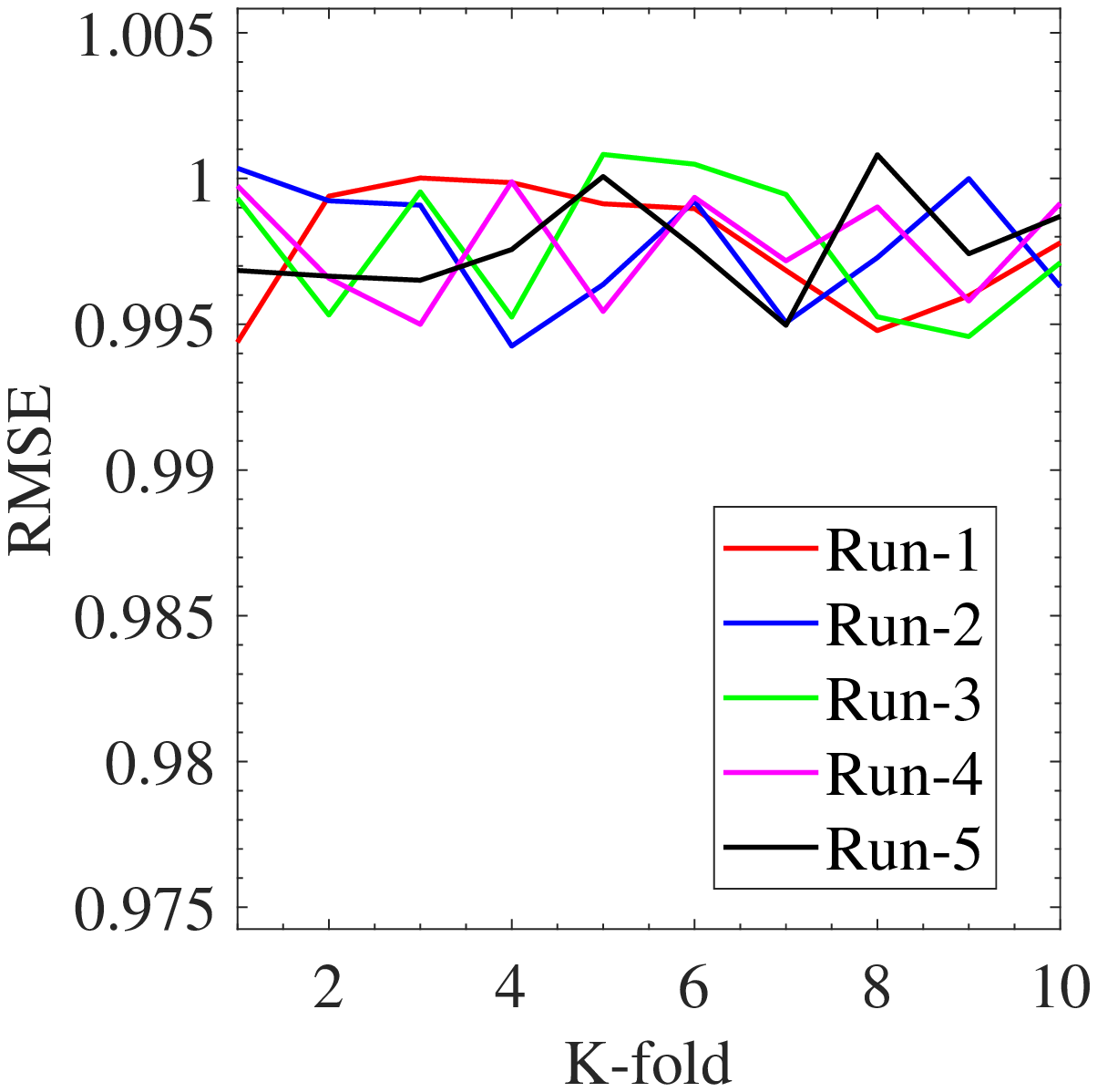}
    \caption{RMSE}
  \end{subfigure}
  \hfill
  \begin{subfigure}[t]{0.32\linewidth}
    \centering
    \includegraphics[width=\linewidth]{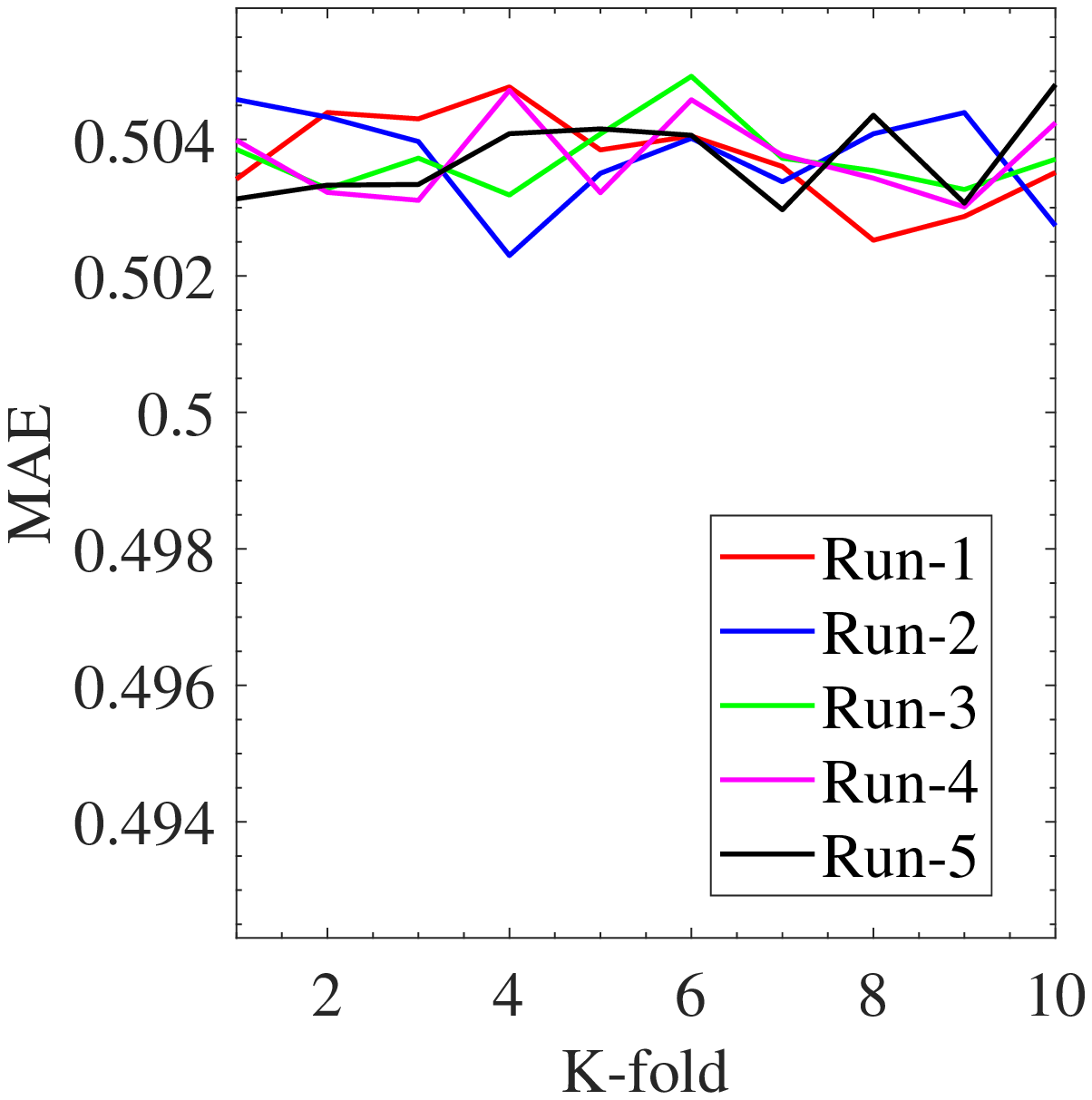}
    \caption{MAE}
  \end{subfigure}
  \hfill
  \begin{subfigure}[t]{0.32\linewidth}
    \centering
    \includegraphics[width=\linewidth]{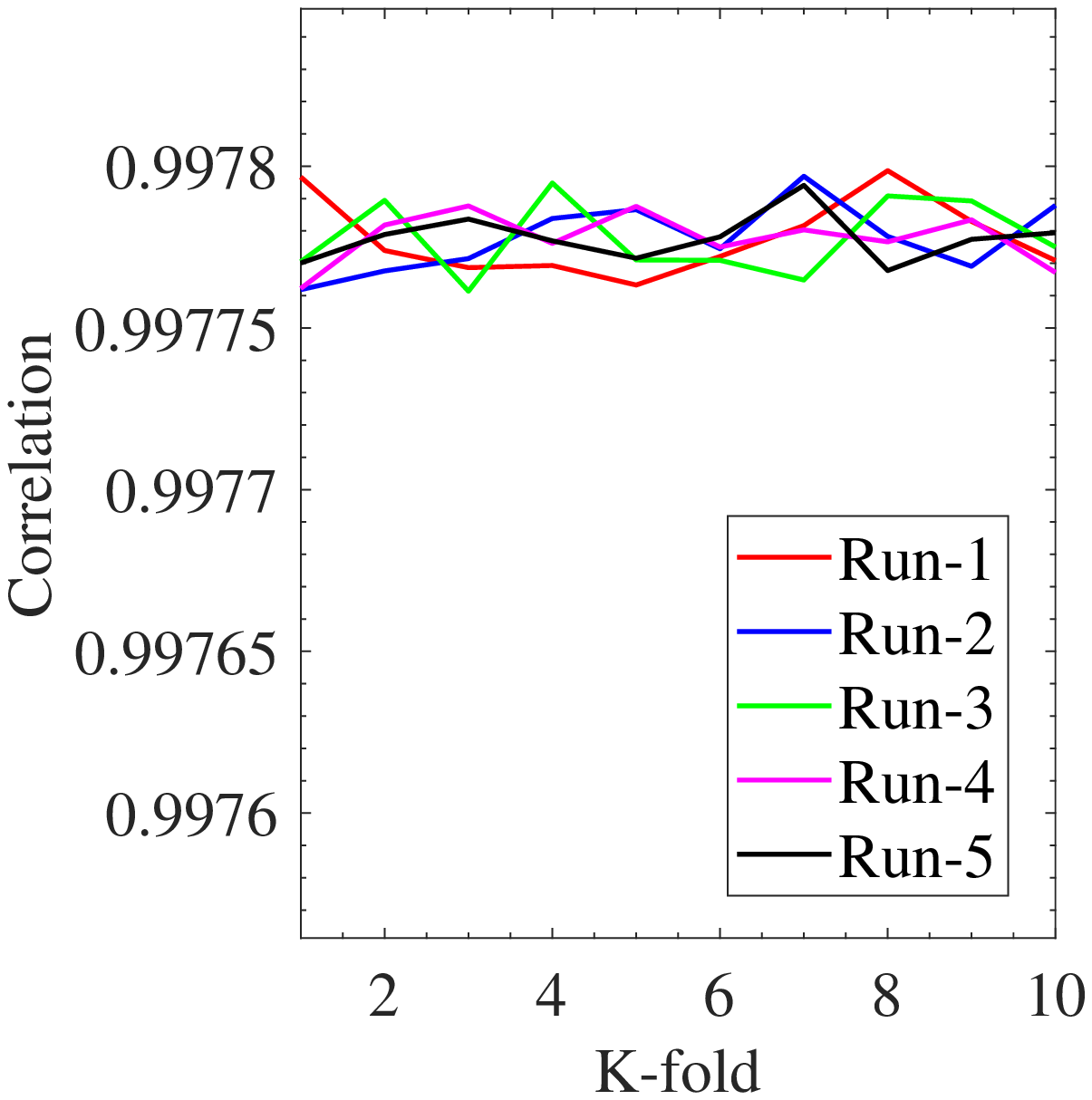}
    \caption{Correlation}
  \end{subfigure}
  
  \caption{Results of repeated 10-fold cross validation.}\label{Fig:CrossValidationMetrics}
\end{figure*}

In order to perform the 10-fold cross validation once for the proposed approach, dataset $DS-I$ was divided into equally sized 10 small datasets. 70\% of these equally sized 10 datasets were selected as the training set and remaining 30\% were used as validation set. Then the CNN model of PCE Module and bagged decision tree of the Fine Tuner Module were trained using the training set and validated using the validation set. The process of selecting the training-validation set and training and validating the proposed approach using the training-validation set was repeated 10 times (the folds). The datasets for training-validation set were selected, such that each of the 10 equally sized datasets is part of the validation set at least once. The process of 10-fold cross validation was repeated 5 times (the runs). The performance of the proposed approach during the repeated 10-fold cross validation has been shown in Figure \ref{Fig:CrossValidationMetrics}. From the figure, it can be observed that the results from different runs overlap and also the results does not vary much. So, it can be concluded that the proposed approach performed consistently well during each run of repeated 10-fold cross validation with low variance.

\subsection{Comparative Analysis}\label{SubSec:ComparativeAnalysis}
The results of four existing approaches, presented in \cite{GALVIN2017234,YANG201441,6861542,MODI2019}, are compared with the proposed approach to benchmark the results. The existing techniques were implemented for comparison as discussed below:

\begin{enumerate}[i)]
    \item Galvin \cite{GALVIN2017234} proposed a multivariate model for estimation of EVs power consumption considering the speed and acceleration of the vehicle as input parameters. Following equation represent his model for NissanSV:

        \begin{equation}\label{Eq:GalvinNissanSVModel}
            P = 479.1 V - 18.93 V^2 + 0.7876 V^3 + 1507 VA
        \end{equation}

        where $P$, $A$ and $V$ is the power consumption, acceleration and speed of the EV, respectively.

    \item Yang et al. \cite{YANG201441} developed a model to estimate power consumption also taking into account the regenerative mode of the motor operation. They gave two Eqs. \eqref{Eq:YangPowerConsumption} and \eqref{Eq:YangRegenrativePower} for normal operation mode and regenerative operation mode, respectively.

        \begin{equation}\label{Eq:YangPowerConsumption}
            P = \frac{v}{\eta_{te}\eta_e}\bigg(\delta m \frac{dv}{dt} + mg(f+i) + \frac{\rho C_D A}{2}v^2\bigg) + P_{accessory}
        \end{equation}

        \begin{equation}\label{Eq:YangRegenrativePower}
            P_{reg} = kv\eta_{te}\eta_m\bigg(\delta m \frac{dv}{dt} + mg(f+i) + \frac{\rho C_D A}{2}v^2\bigg) + P_{accessory}
        \end{equation}

        where $P$, $P_{reg}$, $v$, $\eta_{te}$, $\delta$, $m$, $f$, $i$, $\rho$, $C_D$, $A$, $P_{accessory}$, $k$ and $\eta_m$ represent the power consumption, power regenerated, vehicle speed, transmission efficiency, driving efficiency, EV's weight related coefficient, EV's mass, rolling resistance coefficient, road grade, air density, aerodynamic drag coefficient, frontal area of vehicle, power consumed by accessories, percentage of total energy which can be restored by the motor during braking and motor efficiency, respectively. Parameter $k$ has been defined based on the vehicle's speed as in Eq. \ref{Eq:ParameterK}:

        \begin{equation}\label{Eq:ParameterK}
                k =
                        \begin{dcases}
                            0.5+0.3\frac{v-5}{20}   &  v \geq 5\text{m/s} \\
                            0.5*\frac{v}{5}         &  v < 5\text{m/s}
                        \end{dcases}
        \end{equation}

        The above model has been implemented using the values of $m$, $C_D$ and $A$ provided in \cite{factNissan2013, burress2012benchmarking,MODI2019} for Nissan Leaf. The values of $\delta$, $\rho$, $\eta_{te}$, $\eta_m$, $\eta_e$ and $f$ are given in \cite{YANG201441} as 1.1, 1.2, 0.9, 0.9, 0.8 and 0.015, respectively.

    \item Alvarez et al. \cite{6861542} proposed a neural network model with 1 output and 14 inputs but without hidden layer. The 14 inputs are the mean and variance of vehicle speed, positive acceleration, negative acceleration, SMJ (Starting Movement Jerk), SBJ (Starting Brake Jerk), CTJ (Cruising Track Jerk) and EBJ (Ending Brake Jerk). The output of the network is the energy consumed by the vehicle for the complete trip. The neural network was implemented by training it for 70\% of the data from the dataset $DS-I$ and validated using the remaining 30\%.

    \item A deep learning based CNN model proposed by Modi et al. \cite{MODI2019} was used for comparing the results of the proposed technique. The CNN model proposed by Modi et al. takes three inputs namely, road elevation, vehicle speed and tractive effort and give power consumption as the output. The tractive effort was calculated using the Eq. \eqref{Eq:TractiveEffort}:

        \begin{equation}\label{Eq:TractiveEffort}
            t_{eff} = f_{ad} + f_{rr} + f_{hc} + f_{la} + f_{wa}
        \end{equation}

        where $t_{eff}$, $f_{ad}$, $f_{rr}$, $f_{hc}$, $f_{la}$ and $f_{wa}$ represent the tractive effort, the force required to overcome aerodynamic drag, the rolling resistance force, the gravitational force which plays its role when road elevation changes, the opposing force because of linear acceleration and the inertial force acting on the rotating parts of the vehicle. The CNN model was trained using the 70\% of the data from the dataset $DS-I$ and validated using the rest.
\end{enumerate}

\begin{figure}[h!]
  \centering
  \begin{subfigure}[]{0.49\textwidth}
    \centering
    \includegraphics[width=\textwidth]{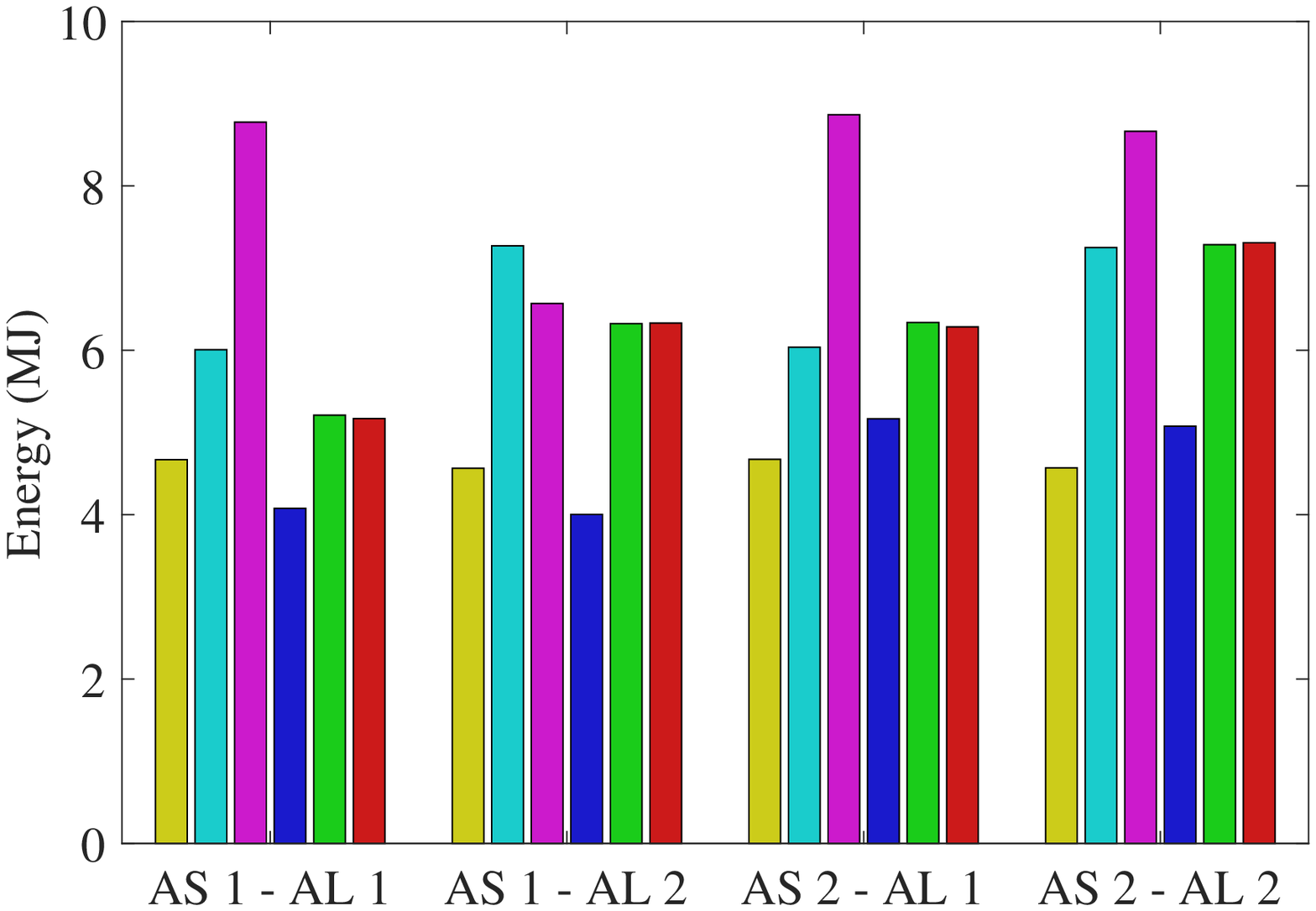}
    \caption{Results with grade profile 1 and temperature $-5^o C$}\label{SubFig:comparison_G0C30T-5}
  \end{subfigure}
  \hfill
  \begin{subfigure}[]{0.49\textwidth}
    \centering
    \includegraphics[width=\linewidth]{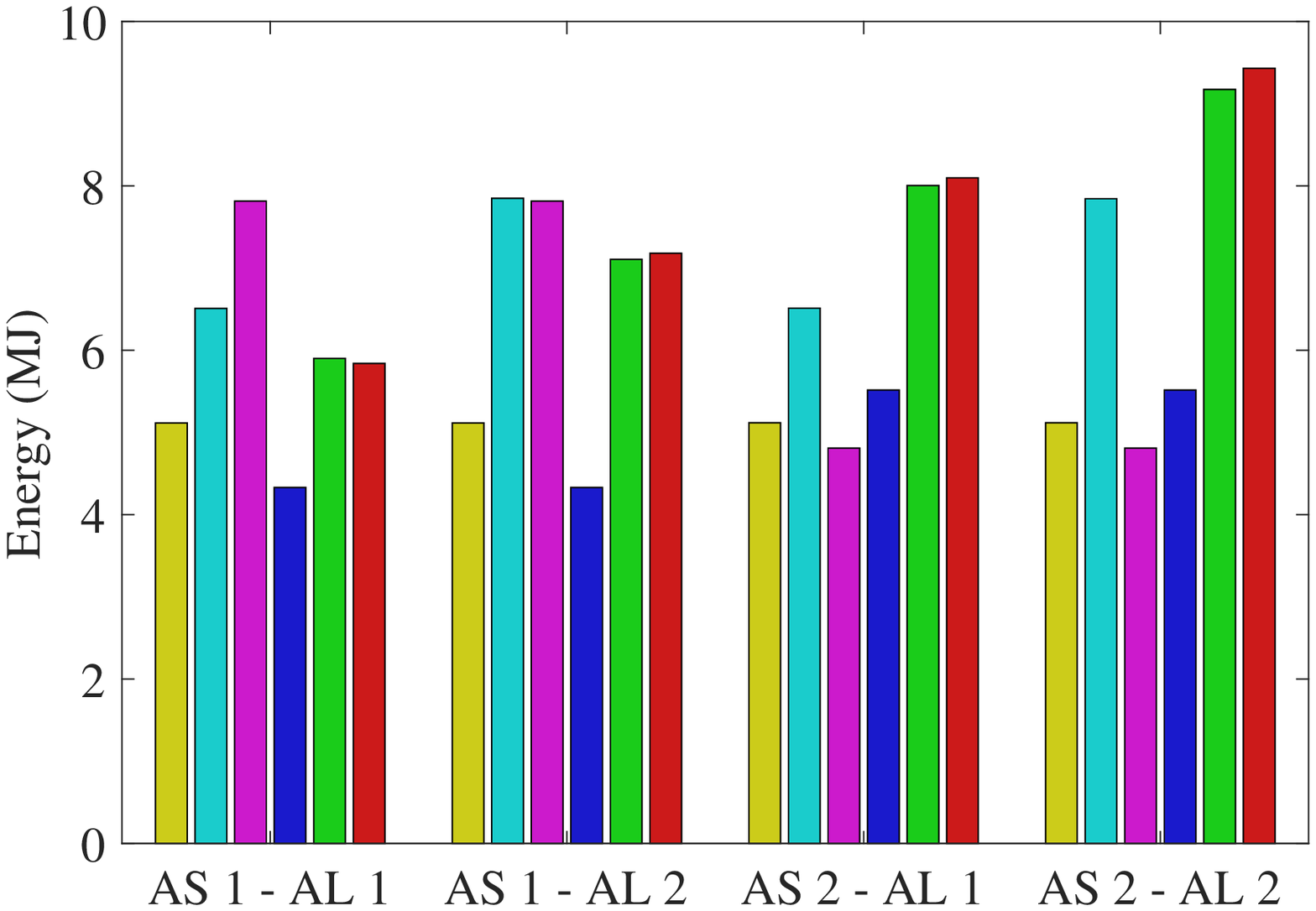}
    \caption{Results with grade profile 1 and temperature $35^o C$}\label{SubFig:comparison_G0C30T35}
  \end{subfigure}

  \begin{subfigure}[]{0.49\textwidth}
    \centering
    \includegraphics[width=\linewidth]{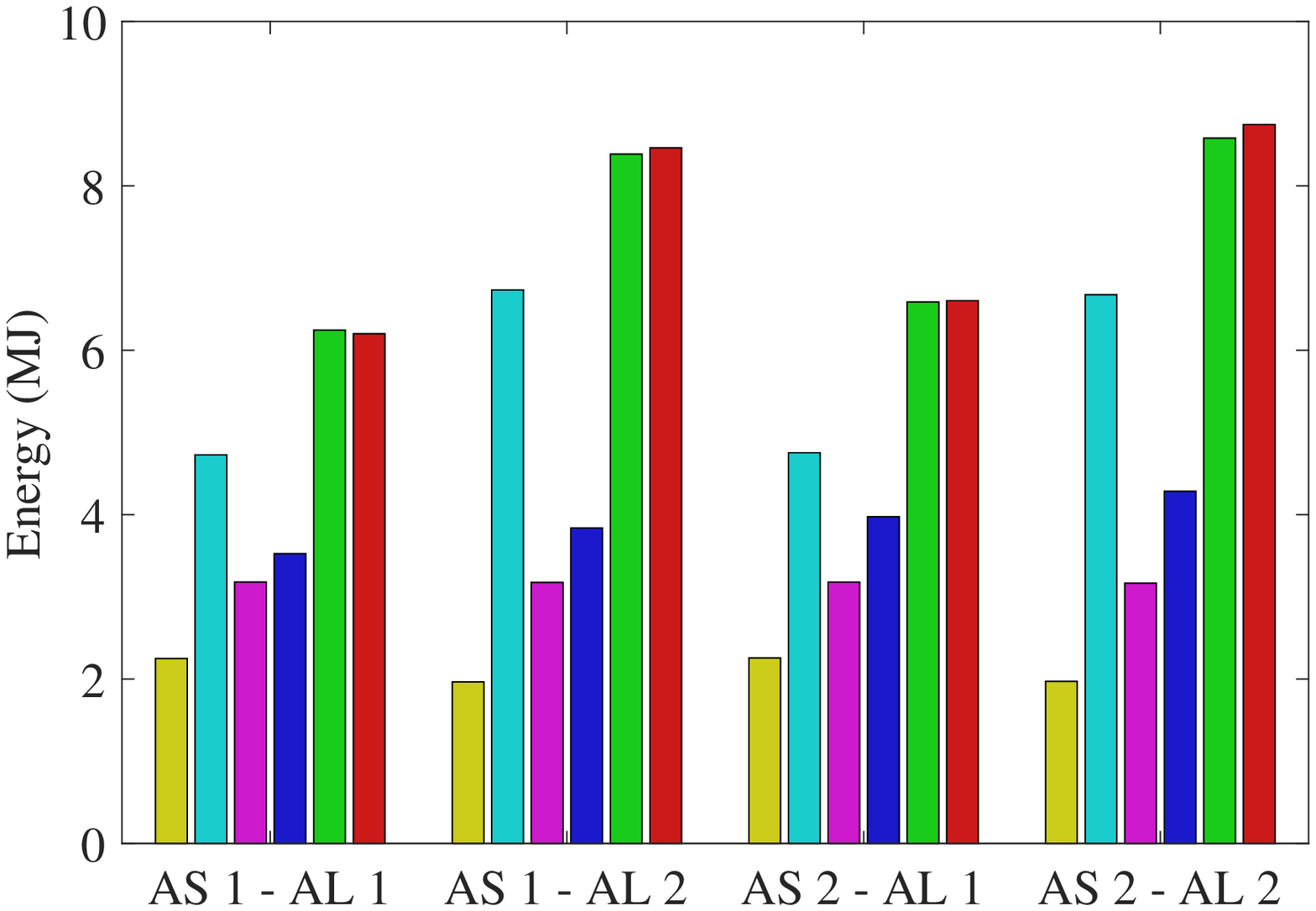}
    \caption{Results with grade profile 2 and temperature $-5^o C$}\label{SubFig:comparison_G10C30T-5}
  \end{subfigure}
  \hfill
  \begin{subfigure}[]{0.49\textwidth}
    \centering
    \includegraphics[width=\linewidth]{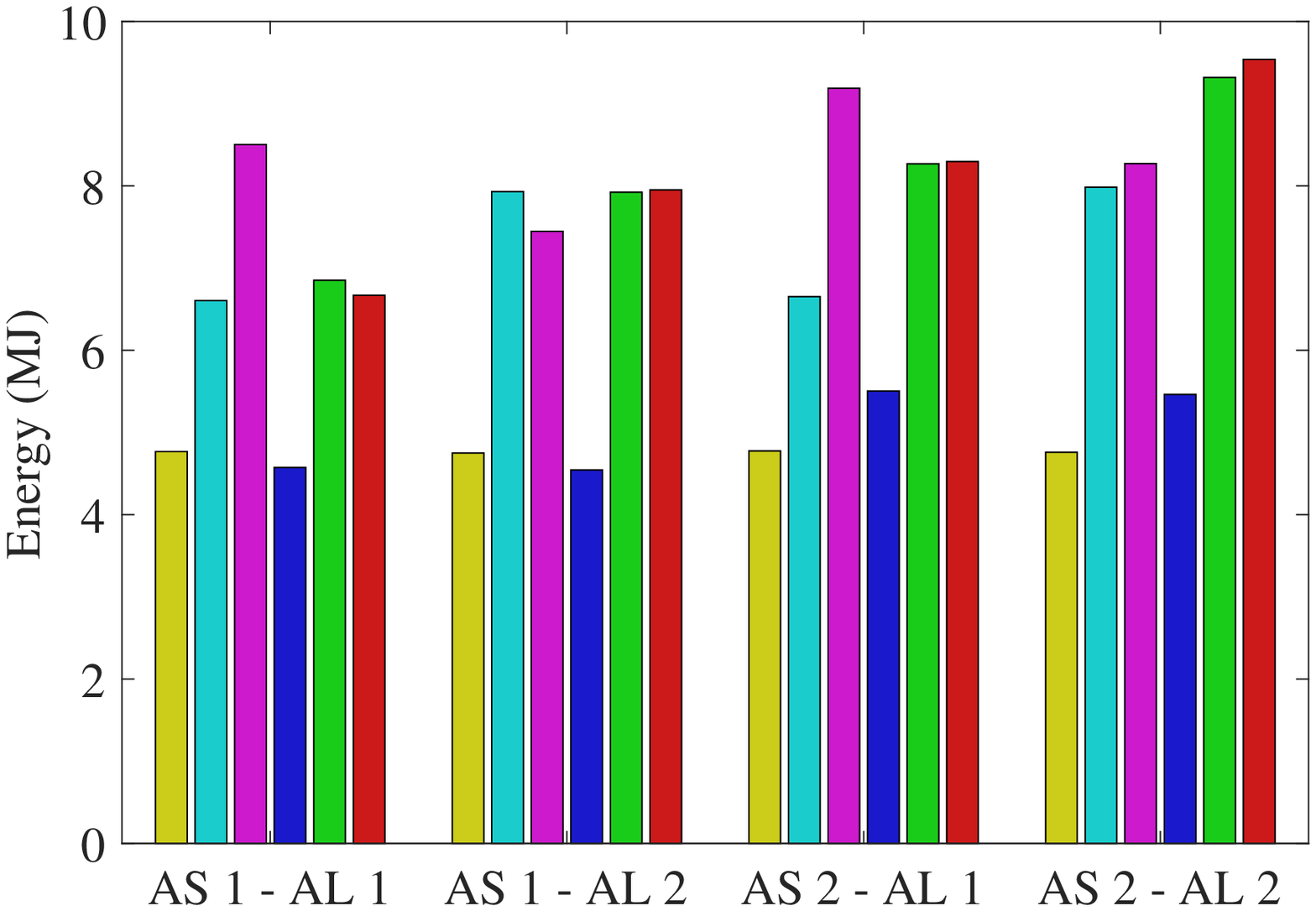}
    \caption{Results with grade profile 2 and temperature $35^o C$}\label{SubFig:comparison_G10C30T35}
  \end{subfigure}

  \caption{Energy consumption prediction comparison of the proposed methodology with other existing techniques for UDDS drive cycle at initial SOC of 30\% under different conditions of two road grade profile, two environmental temperatures, two air speed profiles (AS 1 and AS 2) and two auxiliary loads profiles (AL 1 and AL 2). Legend: Energy consumption: (\protect\solidsquare[henna]) Galvin \cite{GALVIN2017234}, (\protect\solidsquare[skyblue]) Yang et al. \cite{YANG201441}, (\protect\solidsquare[purple]) Alvarez et al. \cite{6861542}, (\protect\solidsquare[darkblue]) Modi et al. \cite{MODI2019}, (\protect\solidsquare[darkgreen]) Proposed Approach, (\protect\solidsquare[darkred]) Actual.} \label{Fig:ComparisonResultsAtSOC30}
\end{figure}

\begin{figure}[h!]
  \centering
  \begin{subfigure}[]{0.49\textwidth}
    \centering
    \includegraphics[width=\textwidth]{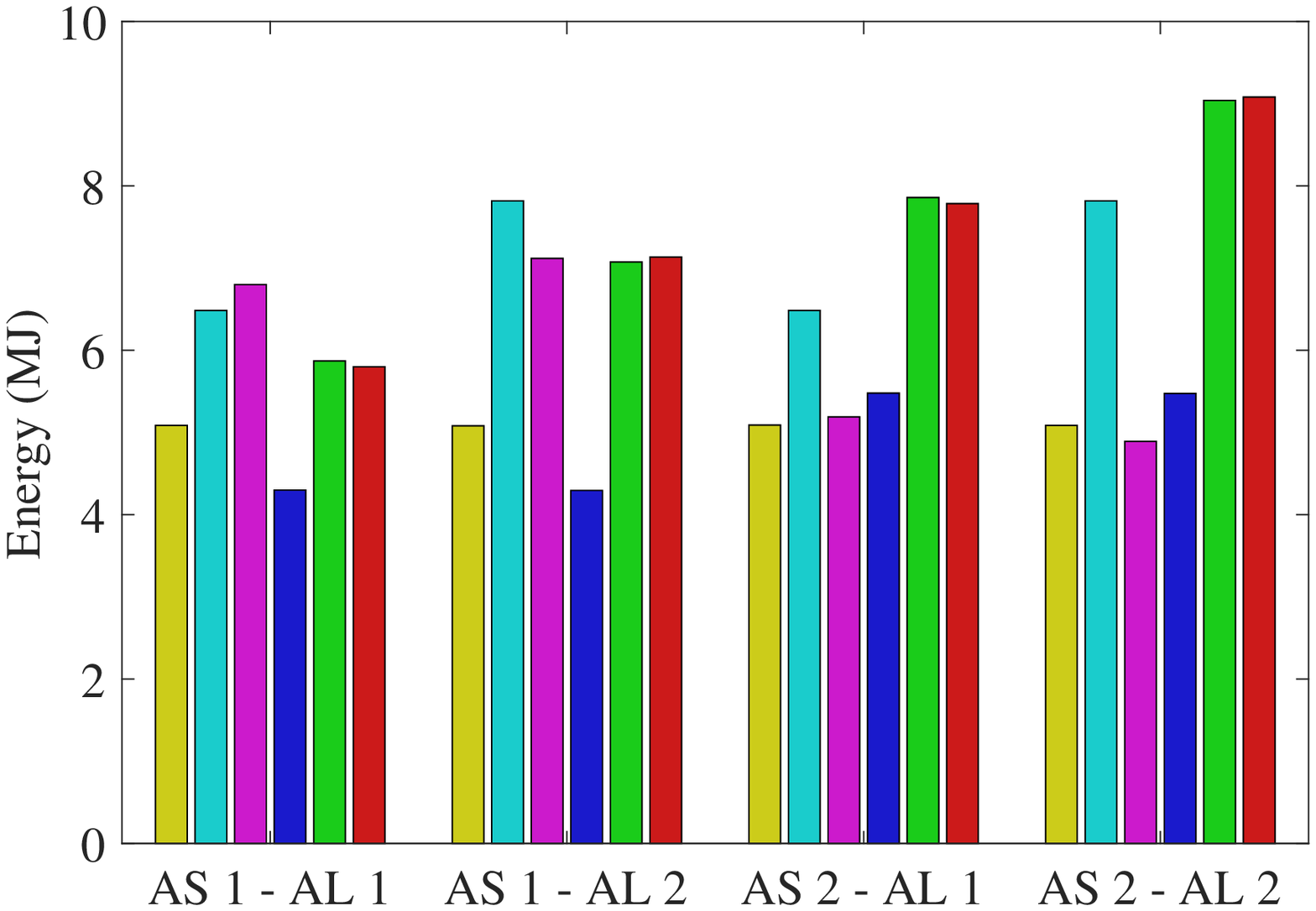}
    \caption{Results with grade profile 1 and temperature $-5^o C$}\label{SubFig:comparison_G0C70T-5}
  \end{subfigure}
  \hfill
  \begin{subfigure}[]{0.49\textwidth}
    \centering
    \includegraphics[width=\linewidth]{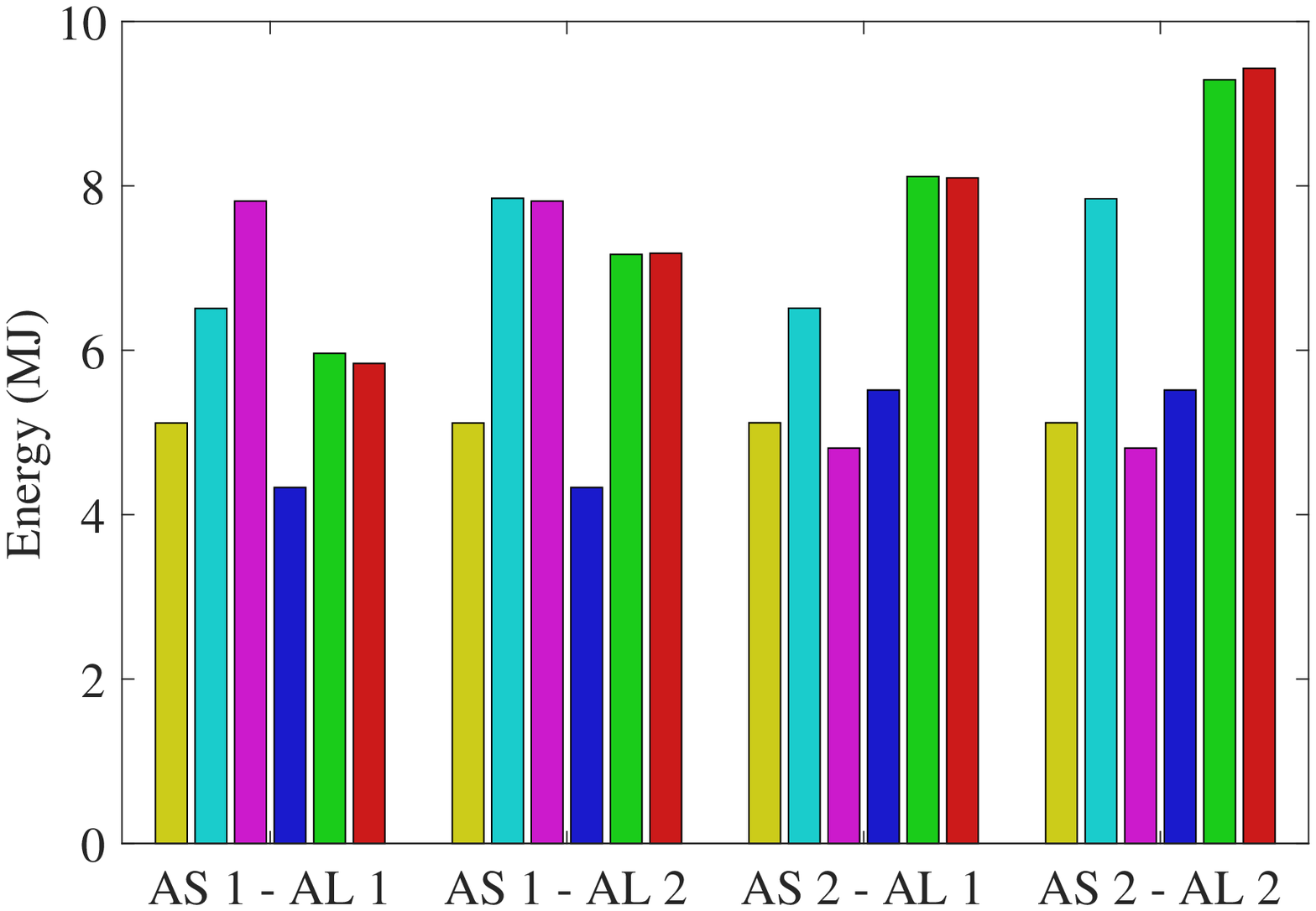}
    \caption{Results with grade profile 1 and temperature $35^o C$}\label{SubFig:comparison_G0C70T35}
  \end{subfigure}

  \begin{subfigure}[]{0.49\textwidth}
    \centering
    \includegraphics[width=\linewidth]{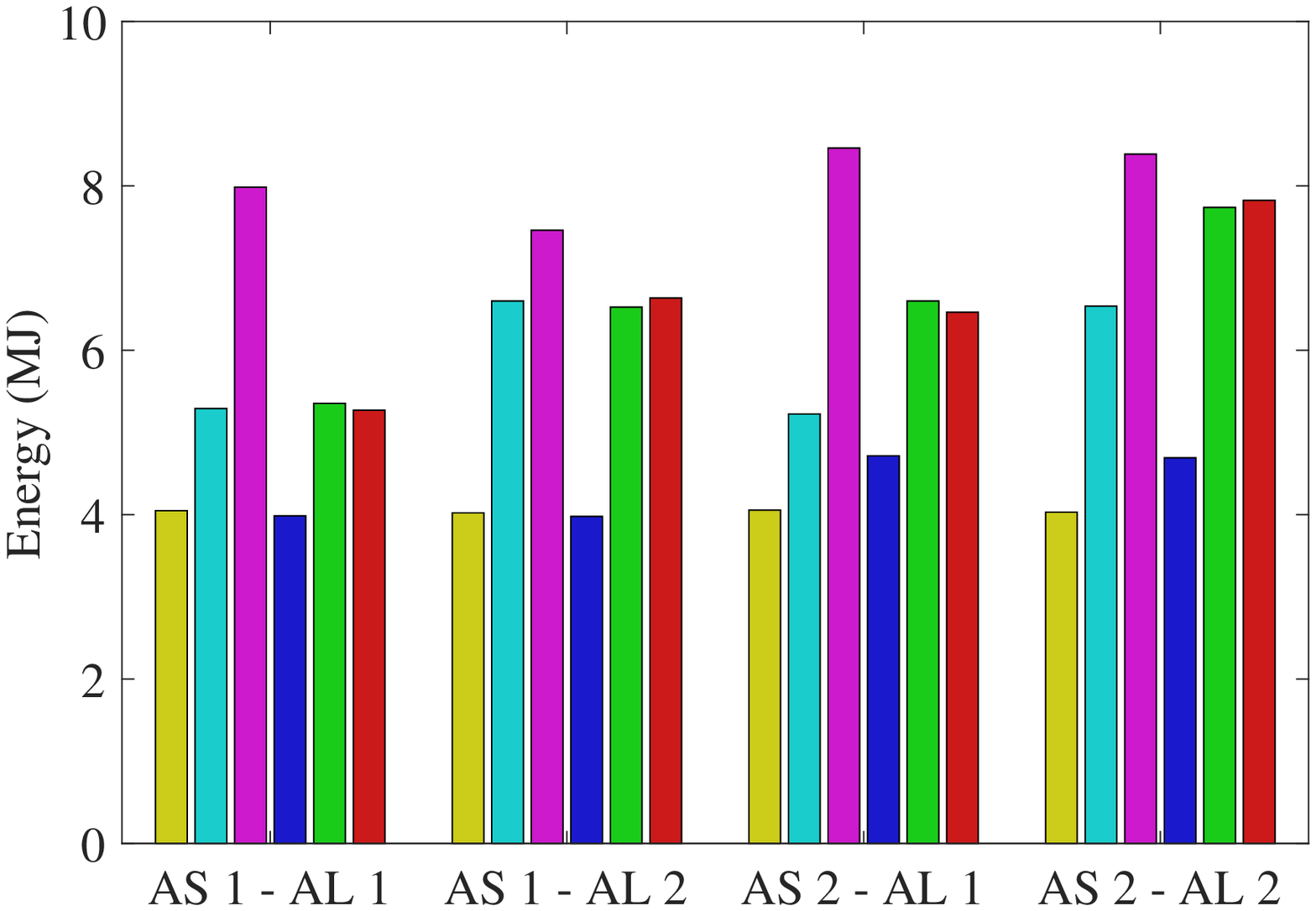}
    \caption{Results with grade profile 2 and temperature $-5^o C$}\label{SubFig:comparison_G10C70T-5}
  \end{subfigure}
  \hfill
  \begin{subfigure}[]{0.49\textwidth}
    \centering
    \includegraphics[width=\linewidth]{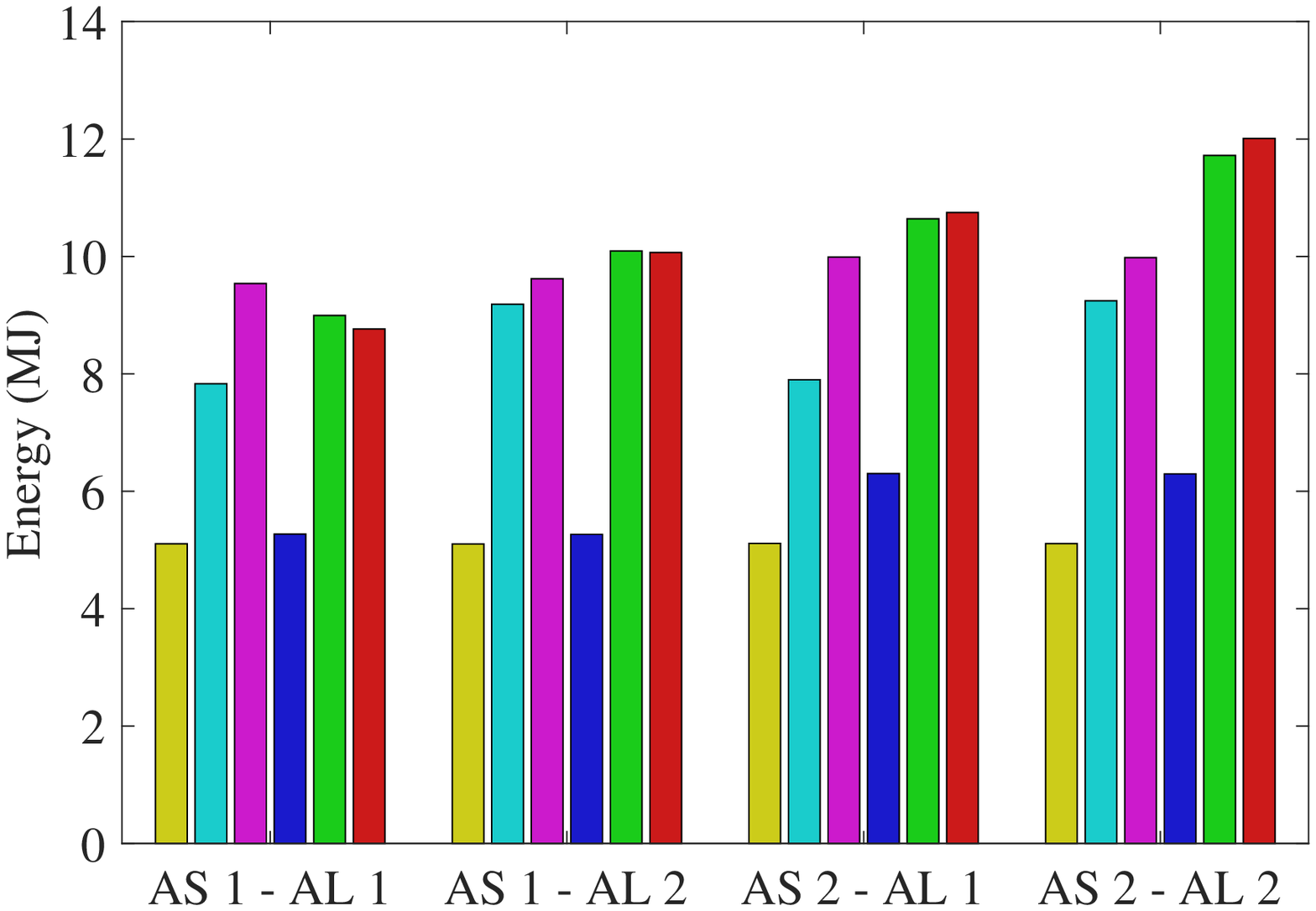}
    \caption{Results with grade profile 2 and temperature $35^o C$}\label{SubFig:comparison_G10C70T35}
  \end{subfigure}

  \caption{Energy consumption prediction comparison of the proposed methodology with other existing techniques for UDDS drive cycle at initial SOC of 70\% under different conditions of two road grade profile, two environmental temperatures, two air speed profiles (AS 1 and AS 2) and two auxiliary loads profiles (AL 1 and AL 2). Legend: Energy consumption: (\protect\solidsquare[henna]) Galvin \cite{GALVIN2017234}, (\protect\solidsquare[skyblue]) Yang et al. \cite{YANG201441}, (\protect\solidsquare[purple]) Alvarez et al. \cite{6861542}, (\protect\solidsquare[darkblue]) Modi et al. \cite{MODI2019}, (\protect\solidsquare[darkgreen]) Proposed Approach, (\protect\solidsquare[darkred]) Actual.} \label{Fig:ComparisonResultsAtSOC70}
\end{figure}

A comparison of estimated energy consumption for UDDS drive cycle under different conditions of the proposed approach and four of the existing techniques, discussed above, has been presented in the Figures \ref{Fig:ComparisonResultsAtSOC30} and \ref{Fig:ComparisonResultsAtSOC70}. Figure \ref{Fig:ComparisonResultsAtSOC30} show the comparison result when the initial SOC of battery was 30\% whereas Figure \ref{Fig:ComparisonResultsAtSOC70} show the results with initial SOC of 70\%. In both the figures there are two rows and two columns, where each row depicts the results at different grade profile and each column show the results at different environment temperature. Each subfigure show the comparison results with four combination of two different air speed profiles (AS 1 and AS 2) and two different auxiliary load profiles (AL 1 and AL 2). The two different grade profiles, air speed profiles and auxiliary load profiles used for the comparison results shown in Figures \ref{Fig:ComparisonResultsAtSOC30} and \ref{Fig:ComparisonResultsAtSOC70} are given in Table \ref{Tab:ParameterProfiles}. From the figures, it can be observed that the proposed approach gave consistently better energy consumption estimate than the existing techniques under all the different conditions. As the technique proposed by Galvin \cite{GALVIN2017234}, considered speed and acceleration only to estimate the power/energy consumption it is justifiable for the results to deviate from actual when other factors come into play. Similarly, Yang et al. \cite{YANG201441} in their proposed technique does not consider the effect of environment temperature and initial SOC on energy consumption and hence the results are not accurate but their technique performed much better than Galvin's technique. The neural network, proposed by Alvarez et al. \cite{6861542}, was not able to accurately predict the result because of two main reasons, first it has no hidden layer due to which it was not able to learn the non-linearity among the influencing factors and second only three parameters namely, speed, acceleration and jerk were taken as input and the influence of other factors was not considered. The deep learning based CNN model, proposed by Modi et al. \cite{MODI2019}, gave consistently lower estimates than the actual energy consumption. The main reason for this is that CNN model developed by Modi et al. does not take into account the effect of varying auxiliary load, air speed, environment temperature and initial SOC of battery while estimating EVs energy consumption.

To generalize the comparison results, Table \ref{Tab:MeanEnergyConsumptionComparison} show the results for different performance metrics for the existing and proposed approach. The different performance metrics are Mean Absolute Energy Deviation ($MAE_{dev}$), Root Mean Square Error ($RMSE$), Mean Absolute Error ($MAE$), Correlation ($Corr$) and Mean Prediction Time per Drive Cycle ($MPTDC$). $MAE_{dev}$ can be obtained using the following equation:

\begin{equation}
MAE_{dev} = \frac{\sum_{i=1}^n \mid E_{act}^i - E_{est}^i \mid}{n}
\end{equation}

where $E_{act}^i$ and $E_{est}^i$ represent the actual and estimated energy consumption for $i^{th}$ drive cycle and $n$ represent total number of drive cycles under consideration. $RMSE$, $MAE$ and $Corr$ are same as defined in Section \ref{SubSec:CrossValidation}. $MPTDC$ is the average time an approach takes to predict the power consumption by the EV for a given drive cycle. It does not include the training time. It can be observed from the Table \ref{Tab:MeanEnergyConsumptionComparison} that the proposed approach gave very reliable results than the existing techniques. The proposed approach has lowest $MAE_{dev}$, $RMSE$, $MAE$ of 0.14, 0.97 and 0.50, respectively and highest $Corr$ of 0.997 for dataset $DS-I_{val}$ in comparison to the other existing techniques. Similar behaviour can be observed for dataset $DS-II$. As dataset $DS-II$ contains data recorded for road grade of $0\%$ and no external wind, due to this all the approaches performed better on $DS-II$ as compared to $DS-I_{val}$. The values of $MAE$, $RMSE$ and $Corr$ are not available for \cite{6861542}, as the NN proposed in it provide total energy consumed as output for the whole trip instead of providing instantaneous power consumption. Due to this, using NN model it is difficult to provide instructions in real-time to the driver based on the current power consumption. Also, there are number of influencing parameters (like initial SOC of battery, wind speed and environmental temperature etc.) that need to be considered along with the parameters considered in \cite{GALVIN2017234,YANG201441,6861542,MODI2019}. Due to this, the proposed technique gave reliable estimates even when these parameters come into play. The proposed technique provides very good results but it takes more time as compared to the existing techniques which can be seen by the values of the fifth metric $MPTDC$, provided in the Table \ref{Tab:MeanEnergyConsumptionComparison}. The proposed technique takes more time to predict results because it require a lot of computation in the PCE module and Fine Tuner module. However, the objective of this work is to develop a technique which can provide accurate results in comparison to other existing techniques. As the results presented in Table \ref{Tab:MeanEnergyConsumptionComparison}, are computed using system with 8 GB RAM and Intel i5 Processor, the real time performance can be achieved by converting the proposed technique into TensorFlow Lite format and execute on Odroid boards with Movidius sticks or Google Coral boards.

\begin{table}[h!]
    \caption{Comparison with existing techniques using different performance metrics}
    \label{Tab:MeanEnergyConsumptionComparison}
    \resizebox{\linewidth}{!}{

    \begin{tabular}{|c|c|c|c|c|c|c|c|c|c|}
        \hline
        \multirow{2}{*}{Approach} & \multicolumn{2}{c|}{$MAE_{dev}$} &  \multicolumn{2}{c|}{$RMSE$} & \multicolumn{2}{c|}{$MAE$} &  \multicolumn{2}{c|}{$Corr$} & $MPTDC$\\ \cline{2-9}

        & $DS-I_{val}$ & $DS-II$ & $DS-I_{val}$ & $DS-II$ & $DS-I_{val}$ & $DS-II$ & $DS-I_{val}$ & $DS-II$ & (in sec)\\ \hline


        Yang et al. \cite{YANG201441} & 2.30 & 1.98 & 8.90  & 3.78 & 4.73 & 2.28 & 0.884 & 0.961 & 1.97 $\times 10^{-3}$ \\ \hline

        Galvin \cite{GALVIN2017234}   & 6.77 & 2.75 & 13.79 & 4.04 & 8.64 & 2.64 & 0.377 & 0.970 & \textbf{3.47} $\mathbf{\times} \mathbf{10^{-4}}$ \\ \hline

        Alvarez et al. \cite{6861542} & 4.68 & 2.15 & NA    & NA   & NA   & NA   & NA    & NA    & 1.14 $\times 10^{-2}$ \\ \hline

        Modi et al. \cite{MODI2019}   & 4.15 & 1.82 & 7.98  & 2.61 & 5.33 & 1.96 & 0.948 & 0.977 & 1.76                  \\ \hline

        Proposed CNN-BDT Model & \textbf{0.14} & \textbf{0.08} & \textbf{0.97} & \textbf{0.74} &  \textbf{0.50} & \textbf{0.41} &  \textbf{0.997} & \textbf{0.998} &  2.10 $\times 10^{-1}$    \\ \hline

        \multicolumn{10}{|l|}{\footnotesize{Values in bold represent the best ones and NA means not applicable}}     \\ \hline
    \end{tabular}

    }
\end{table}

\subsection{Statistical Analysis}\label{SubSec:StatisticalAnalysis}
To further validate the conclusion that the proposed approach is better than existing techniques, statistical analysis using two sample $t$-test has also been performed between the results of proposed approach and other existing techniques i.e. proposed approach with Galvin \cite{GALVIN2017234}, proposed approach with Yang et al. \cite{YANG201441} and so on. For this, the 10 partitions used for validating the proposed approach during 10-cross validation were used and results were obtained for these partitions using existing techniques. So, for each existing technique $MAE_{dev}$, $RMSE$ and other performance metrics were calculated for the results for each of the 10 partitions. Using these observations an analysis was performed by taking an assumption that with the significance level of $\alpha=0.05$ the populations have equal variances. Population means difference as zero was taken as the null hypothesis. Under the above null hypothesis, two sample $t$-test was performed and Table \ref{Tab:pValueforCNNModels} shows the results obtained. From the table, it can be seen that the values of t-critical are less than the values of $t$-stat for each pair of techniques and also the p-values are less than the significance level $\alpha=0.05$. This implies that means of the populations differ significantly and hence null hypothesis is rejected. As the mean of the $MAE_{dev}$ and $RMSE$ values is less for the proposed approach than other existing techniques thus, it can be concluded that the proposed approach provide better results and the difference in results is statistically significant.

\begin{table}[h!]
    \centering

    \caption{Results of two sample $t$-test for statistical analysis}
    \label{Tab:pValueforCNNModels}
    \resizebox{\linewidth}{!}{
    \begin{tabular}{|c|c|c|c|c|c|}
    \hline
                    &Galvin \cite{GALVIN2017234} & Yang et al. \cite{YANG201441} & Alvarez et al. \cite{6861542} & Modi et al. \cite{MODI2019} & Proposed Approach                                                  \\ \hline

                    &\multicolumn{5}{c|}{Results of $t$-test on $MAE_{dev}$ from 10-fold cross validation} \\ \hline
    Observations					 & 10     & 10     & 10     & 10     & 10                  \\ \hline
    Mean 						     & 6.772  & 2.306  & 4.679  & 4.155  & 0.139               \\ \hline
    Variance						 & 29.660 & 4.104  & 16.258 & 4.710  & 0.012               \\ \hline
    Hypothetical mean difference	 & 0      & 0      & 0      & 0      & -                   \\ \hline
    Pooled Variance					 & 15.045 & 2.087  & 8.249  & 2.394  & -                   \\ \hline
    Degree of freedom				 & 18     & 18     & 18     & 18     & -                   \\ \hline
    t–critical one tail		         & 1.734  & 1.734  & 1.734  & 1.734  & -                   \\ \hline
    $P(T \leq t)$ one tail			 & 0      & 0      & 0      & 0      & -                   \\ \hline
    $t$-stat						 & 10.261 & 9.002  & 9.485  & 15.572 & -                   \\ \hline

                    &\multicolumn{5}{c|}{Results of $t$-test on $RMSE$ from 10-fold cross validation}      \\ \hline
    Observations					 & 10     & 10     & -      & 10     & 10                  \\ \hline
    Mean 						     & 13.796 & 8.900  & -      & 7.988  & 0.975               \\ \hline
    Variance						 & 37.372 & 24.303 & -      & 1.255  & 0.047               \\ \hline
    Hypothetical mean difference	 & 0      & 0      & -      & 0      & -                   \\ \hline
    Pooled Variance					 & 20.788 & 13.528 & -      & 0.724  & -                   \\ \hline
    Degree of freedom				 & 18     & 18     & -      & 18     & -                   \\ \hline
    t–critical one tail		         & 1.734  & 1.734  & -      & 1.734  & -                   \\ \hline
    $P(T \leq t)$ one tail			 & 3.159 $\times 10^{-6}$ & 6.902 $\times 10^{-5}$ & - & 0 & -                \\ \hline
    $t$-stat						 & 6.287  & 4.817  & -      & 18.433 & -                   \\ \hline

\end{tabular}}

\end{table}

\section{Conclusion}\label{Sec:Conclusion}
In this paper, a hybrid approach using CNN and BDT has been developed to provide the EV drivers with accurate energy consumption estimates in real-time. The proposed approach consider the effect of seven factors namely, vehicle speed, acceleration, air speed, road elevation, auxiliary loads, environment temperature and initial battery's SOC. Detailed elaboration of developing, training and testing the proposed approach has been provided. The performance of the approach has been validated by comparing it with number of state-of-the-art techniques. Following points include the main conclusions:

\begin{enumerate}[i)]
    \item The proposed CNN-BDT approach can provide energy consumption estimates in real-time. Hence, it can be used to guide the driver in real-time and decrease his/her range anxiety.
    \item The proposed approach does not require any vehicle specific parameters, like battery's internal resistance, motor's efficiency curve, battery's descharging/charging curve etc., for estimating the energy consumption. Due to this, the proposed approach can be easily generalized for any other electric vehicle.
    \item The accuracy of the proposed approach has been validated by comparing the proposed approach with other existing techniques. The comparison results show that proposed approach can estimate the energy consumption with least mean absolute energy deviation of 0.14 and highest correlation of 0.997.
    \item From the comparison results, it can be concluded that the proposed approach, unlike the previous existing techniques, can learn the non-linear relationship of different parameters quite accurately.
    \item The CNN-BDT model can be converted to TensorFlow Lite format and then can easily run on different microcontrollers or can be deployed to vehicular embedded system.
\end{enumerate}
\section*{Compliance with ethical standards}

\textbf{Conflict of interest} All authors declare that they have no conflict of interests.

\textbf{Ethical approval} This article does not contain any studies with human or animals performed by any of the authors.


\end{document}